\RequirePackage{lineno} 

\documentclass[twocolumn,showpacs,preprintnumbers,amsmath,amssymb,nofootinbib]{revtex4}

\usepackage{graphicx}
\usepackage{dcolumn}
\usepackage{bm}

 \usepackage{epstopdf}
\def\bea{\begin{eqnarray}}
\def\eea{\end{eqnarray}}

\def\pp{\mbox{$p$-$p$}}
\def\auau{\mbox{Au-Au}}

\def\pbpb{\mbox{Pb-Pb}}
\def\aa{\mbox{A-A}}
\def\nn{\mbox{N-N}}

\def\ep{\mbox{e-p}}
\def\ee{\mbox{$e^+$-$e^-$}}
\def\ppbar{\mbox{$p$-$\bar p$}}

\def\pt{$p_t$}
\def\mt{$m_t$}
\def\yt{$y_t$}

\def\et{$E_t$}
\def\met{$\langle E_t \rangle$}
\def\meet{$\langle e_t \rangle$}
\def\mpt{$\langle p_t \rangle$}
\def\ppsec{$\bf p$-$\bf p$}
\begin{document} 

\setpagewiselinenumbers
\modulolinenumbers[5]

\preprint{Version 0.7}

\title{Can minimum-bias distributions on transverse energy test hadron production models?
}

\author{Thomas A.\ Trainor}\affiliation{CENPA 354290, University of Washington, Seattle, WA 98195}


\date{\today}

\begin{abstract}
A recent study  reports measurements of transverse-energy $E_t$ distributions at mid-rapidity for several high-energy nuclear collision systems. The $E_t$ data are analyzed in the context of constituent-quark (CQ) participants estimated with a Glauber-model simulation. The study concludes that systematic variations of hadron and \et\ yields previously interpreted in terms of a two-component soft+hard model (TCM) of hadron production including a dijet (hard) contribution are actually the result of CQ participant trends with only soft production. It is claimed that deviations from linear scaling with the number of nucleon participants of hadron yields vs  \aa\ centrality do not actually arise from dijet production as previously assumed. In the present study we examine the new $E_t$ data in the context of the TCM and compare those results with previous differential spectrum and minimum-bias correlation analysis. We present substantial evidence supporting a significant dijet contribution to all high-energy nuclear collisions consistent with the TCM and conclude that the $E_t$ data, given their systematic uncertainties, fail to support claimed CQ model interpretations.
\end{abstract}


\maketitle

 \section{Introduction}
 
Some features of the hadronic final state of high-energy nuclear collisions may carry significant information about  collision dynamics and production mechanisms. In order of successive integrations (and  therefore reduced information) are multiparticle correlations, single-particle spectra, minimum-bias (MB) distributions of event number on integrated particle-number or momentum/energy yields within acceptance windows and centrality distributions of such quantities. This study considers MB distributions within a larger context provided by differential measures. We consider what information can be extracted from MB distributions on transverse energy \et\ and multiplicity $n_{ch}$ within some angular acceptance.
 
 Reference~\cite{phenix} studies mid-rapidity MB distributions on \et\ in the context of soft hadron production and a constituent-quark (CQ) model (CQM). The study concludes that a two-component model (TCM) including soft (projectile dissociation) and hard (jet-related) hadron production conventionally used to describe MB trends on \et\ and $n_{ch}$ serves as a proxy for the real mechanism dominated by soft hadron production from QCD color strings connecting participant CQs, with no significant dijet contribution.
 The arguments supporting that interpretation are based in part on the historical development of hadronic physics since the 1960s and in part on apparent consistency between MB distributions on $n_{ch}$ and \et\ and Monte Carlo simulation of MB distributions on the number of conjectured CQ participants $N_{qp}$~\cite{voloshin}.
 
 The CQ narrative deviates strongly from the body of data  represented by  the TCM as it has developed during the past thirteen years of Relativistic Heavy Ion Collider (RHIC) operation. Study of spectra and correlations from $\sqrt{s_{NN}} = 200$ GeV \pp\ collisions~\cite{ppprd,porter2,porter3} reveals a strong dijet contribution quantitatively consistent with perturbative QCD (pQCD) predictions that combine measured jet spectra describing {Sp\= pS} calorimeter data~\cite{ua1} with measured fragmentation functions (FFs) from the Large Electron-Positron (LEP) collider~\cite{opal,aleph}. The \pp\ dijet systematics can be extrapolated to \aa\ centrality dependence via a Glauber \aa\ geometry model to form a TCM reference. Deviations from the \pp\ extrapolation reference may reveal new physics. In particular, features of the MB distribution on $n_{ch}$ are described in detail by combining the \pp\ TCM with modification of FFs according to a simple pQCD prescription~\cite{borg}.
 
 In this study we review claims that a soft mechanism following CQ scaling describes all mid-rapidity hadron and \et\ production. We summarize experimental evidence from \pp\ and \auau\ data that supports a soft+hard two-component model. We consider the structure of \et\ data from Ref.~\cite{phenix} and find that the reported MB distribution on \et\ for \auau\ collisions is actually consistent with the full TCM, including measured strong centrality variation of a jet-related hard component. We find no necessity to invoke CQ scaling. Certain aspects of \pp\ and \aa\ data appear to falsify that conjecture.
 
 The parameter $x$ as used in this study has two conventional meanings. In the context of QCD hadron structure $x$ is a parton momentum fraction. In the context of the TCM for nuclear collisions $x \approx n_h / n_{pp}$ is the fraction of \pp\ (\nn) multiplicity included in the hard component.

This article is arranged as follows:
Sec.~\ref{ethadron} introduces two-component and CQ models for \et\  and hadron production.
Sec.~\ref{anal} reviews relevant analysis methods for high-energy nuclear collisions.
Sec.~\ref{phetdata} introduces PHENIX \et\  data and CQ interpretations.
Sec.~\ref{ppcoll} summarizes the phenomenology of 200 GeV \pp\ collisions.
Sec.~\ref{aacoll} summarizes the phenomenology of 62.4 and 200 GeV \auau\ collisions.
Sec.~\ref{response} presents a set of challenges for the CQ model based on material in the previous two sections.
Secs.~\ref{disc} and~\ref{summ} present Discussion and Summary.
Three appendices review TCM descriptions of MB distributions, TCM energy dependence and the algebraic structure of joint MB distributions.

\section{$\bf E_t$ and hadron production models} \label{ethadron}

This study compares two-component (significant dijet component) and CQ (no significant dijet component) models for  transverse-energy and hadron production near mid-rapidity. Reference~\cite{phenix} favors a CQM in which hadron production is dominated by soft processes (conjectured fragmentation of color strings connecting CQs) with systematics determined by CQ participant number $N_{qp}$. The role of dijets is assumed to be negligible. Supporting arguments for the CQM are derived from centrality dependence of $E_t$ and $n_{ch}$ production and from the structure of MB distributions on those quantities. The TCM alternative includes dijet production from collisions of small-$x$ gluons as a significant component. Advocates of the CQM assert that the TCM serves only as a proxy for the CQM in describing  \et\ and hadron production. In this section we introduce the two models.

\subsection{The CQM and soft hadron production}

Mid-rapidity measurements of $dn_{ch}/d\eta$ and $dE_t/d\eta$ are reported to give ``excellent characterization of the nuclear geometry...and are sensitive to the underlying reaction dynamics...''~\cite{phenix}. For example, $dn_{ch}/d\eta$ is found to deviate from strict proportionality to the number of participant nucleons $N_{part}$ (linearity) expected for soft hadron production. That deviation has been characterized conventionally by the two-component (soft+hard) model, with dijet production as the hard component. 

However, for the CQM  it is assumed that mid-rapidity hadron and $E_t$ production are dominated by low-\pt\ soft hadrons and should therefore be insensitive to hard processes. An alternative model based on the number of CQ participants $N_{qp}$ includes only soft-hadron production (no dijets), with produced $n_{ch}$ and $E_t$ proportional to the number of color strings that connect constituent quarks. The relation $dn_{ch}/d\eta \propto N_{qp}$ is assume~\cite{voloshin} based on data from Ref.~\cite{phobostcm} that cover only the most-central 40\% of the \auau\ total cross section. 

Reference~\cite{phenix} claims to establish the same proportionality for $dE_t/d\eta$, but the analysis and its interpretation are constrained by critical assumptions: ``...possible models motivated by the fact that half the momentum of a nucleon is carried by [small-$x$] gluons when probed at high $Q^2$ in hard-scattering are not considered...we limit our comparison to the nucleon and [large-$x$] constituent-quark participant models...widely used since the 1970's....'' The study concludes that  ``...the success of the two component model [of hadron and \et\ production] is not because  there are some contributions proportional to $N_{part}$ and some proportional to $N_{coll}$ [$\equiv N_{bin}$], but rather because a particular linear combination of $N_{part}$ and $N_{coll}$ turns out to be an empirical proxy for the nuclear geometry of the number of constituent quark participants, $N_{qp}$ in A+A collisions.'' Thus, the TCM ``...does not represent a hard-scattering component in  $E_t$ distributions.'' Reference~\cite{phenix} rejects any statistically-significant role for dijets in HE nuclear collisions.

\subsection{The TCM and importance of small-$\bf x$ gluons} \label{lowxglue}

A key assumption in the argument of Ref.~\cite{phenix} is exclusion of the role of small-$x$ partons (mainly gluons) in hadron and $E_t$ production at mid-rapidity. However, that assumption contradicts modern QCD theory. The description of hadron small-$x$ structure (parton distribution functions or PDFs) has been greatly refined over the past two decades. It is likely that small-$x$ gluons dominate hadron production in high-energy nuclear collisions, both as sources of ``soft'' hadron production and, via large-angle parton-parton scattering, as sources of ``hard'' dijet production. The latter can be predicted quantitatively via pQCD calculations given the flux of small-$x$ gluons~\cite{fragevo}.

The TCM has had two principal manifestations at RHIC: (a) phenomenological description of \aa\ production with TCM parameter $x$ fitted to more-central \aa\ data~\cite{kn} and
(b) extrapolation (based on the Glauber model of \aa\ collisions) of a detailed  TCM for \pp\ collisions including direct dijet manifestations in spectra and correlations compatible with pQCD calculations. 
The extrapolation from \pp\ provides a {\em Glauber linear superposition} (GLS) reference for the TCM in \aa\ collisions~\cite{porter2,porter3,ppprd}. 
Reference (b) serves as a null hypothesis: \aa\ centrality trends inconsistent with the GLS reference (e.g., requiring a variable  $x$) may indicate novel physics~\cite{hardspec,fragevo}.

The number of constituent-quark participants $N_{qp}$ is determined by a Glauber-model simulation of \aa\ collisions assuming three CQs per nucleon, with cross sections adjusted to be self-consistent. By hypothesis CQs reside near $x = 1/3$ (dressed quarks or valons~\cite{valons}) whereas according to modern QCD the partons most responsible for mid-rapidity hadron production at RHIC energies should reside near $x = 0.01$. Within the same QCD context the small-$x$ structure of hadrons should be universal, not influenced by the configuration of valence quarks.

In this study we present several examples from \pp\ collision systematics that exclude the CQ hypothesis. We summarize detailed relations among yields, spectra and correlations that are describe accurately by a TCM reference derived from \pp\ data quantitatively related to pQCD calculations. The \pp\ reference is then modified in a simple manner, again consistent with pQCD, to describe a broad array of data from more-central \auau\ collisions. We relate the TCM to model and data MB distributions and demonstrate accurate correspondence.

\section{Analysis methods} \label{anal}
 
 We briefly summarize the kinematic variables, spaces and methods required to describe high-energy nuclear collisions, especially related to \aa\ centrality measurement and yield, spectrum and correlation measurements. Additional method descriptions can be found in Refs.~\cite{ppprd,hardspec,porter2,porter3,inverse,axialci,anomalous,davidhq,davidhq2}.
  
 \subsection{Kinematic variables and spaces} \label{kine}
 
 High-energy nuclear collisions are described efficiently within a cylindrical coordinate system $(p_t,\eta,\phi)$, where $p_t$ is the transverse momentum, $\phi$ is the azimuth angle from a reference direction and pseudorapidity $\eta = - \ln [\tan(\theta/2)] \approx \sin(\pi/2 - \theta)$ is a measure of the polar angle, the approximation being valid near $\eta = 0$. A bounded detector angular acceptance is denoted by intervals ($\Delta \eta,\Delta \phi$) in the primary single-particle space $(\eta,\phi)$.
  
 Although scalar momentum $p_t$ is directly measured by particle detectors in this study we prefer to use an alternative measure. To provide better visual access to low-momentum structure and to simplify the description of jet-related spectrum hard components (defined below) we present single-particle (SP) spectra in terms of  transverse rapidity $y_t = \ln[(m_t + p_t)/ m_h]$ with transverse mass $m_t = \sqrt{p_t^2 + m_h^2}$ and rest mass $m_h = m_\pi$ assumed for unidentified hadrons.  The statistical measure \mpt\ (both event-wise and ensemble means) is retained for analysis of spectra~\cite{ppprd,hardspec}, fluctuations~\cite{ptfluct} and correlations~\cite{ptscale,ptedep}.
 
 The main kinematic quantity for this study is transverse energy \et\ as measured by an electromagnetic calorimeter (EMCal) and integrated within some angular acceptance. We adopt the PHENIX notation convention  $dE_t/d\eta \rightarrow \langle E_t \rangle$. We also consider \et\ per hadron \meet\ = \et\ / $n_{ch}$ within some acceptance.
   
 \subsection{\aa\ centrality measures}
 
 The \aa\ centrality evolution of $n_{ch}$, \et\ and other quantities is a major issue for this study. \aa\ centrality is conventionally measured at RHIC by Glauber-simulated participant-nucleon number $N_{part}$.  However, alternative measures offer better access to the more-peripheral centrality variation required to test relevant hypotheses.
 
Minimum-bias distributions of \aa\ cross section $\sigma$ on participant number $N_{part}$ and \nn\ binary collision number $N_{bin}$ are accurately described by power-law trends, leading to simple parametrizations in terms of the fractional cross section $\sigma/\sigma_0$~\cite{powerlaw}
\bea \label{glauber}
(N_{part}/2)^{1/4} &=& 0.5^{1/4}\frac{\sigma }{\sigma_0} + (N_{part,max}/2)^{1/4} \left(1-\frac{\sigma }{\sigma_0}\right)
\nonumber \\ 
N_{bin}^{1/6} &=&  0.5^{1/6}\frac{\sigma }{\sigma_0} + N_{bin,max}^{1/6}\left(1-\frac{\sigma }{\sigma_0}\right),
\eea
with $N_{part,max} = 382$ and $N_{bin,max} = 1136$ for 200 GeV \auau\ collisions. Those parametrizations describe Glauber simulations (with $\sigma_{NN} = 42$ mb) at the percent level (e.g., within 1\% of the 200 GeV $N_{part}$ values in Table V of Ref.~\cite{phenix}). We use the same values for all energies above $\sqrt{s_{NN}} \approx 30$ GeV as purely geometric centrality measures. The preferred centrality measure for data plots is mean participant path length $\nu = 2 N_{bin} / N_{part}$ which provides good visual access to the more-peripheral data required to test the \nn\ linear-superposition hypothesis.

\begin{figure}[h]
  \includegraphics[width=1.65in]{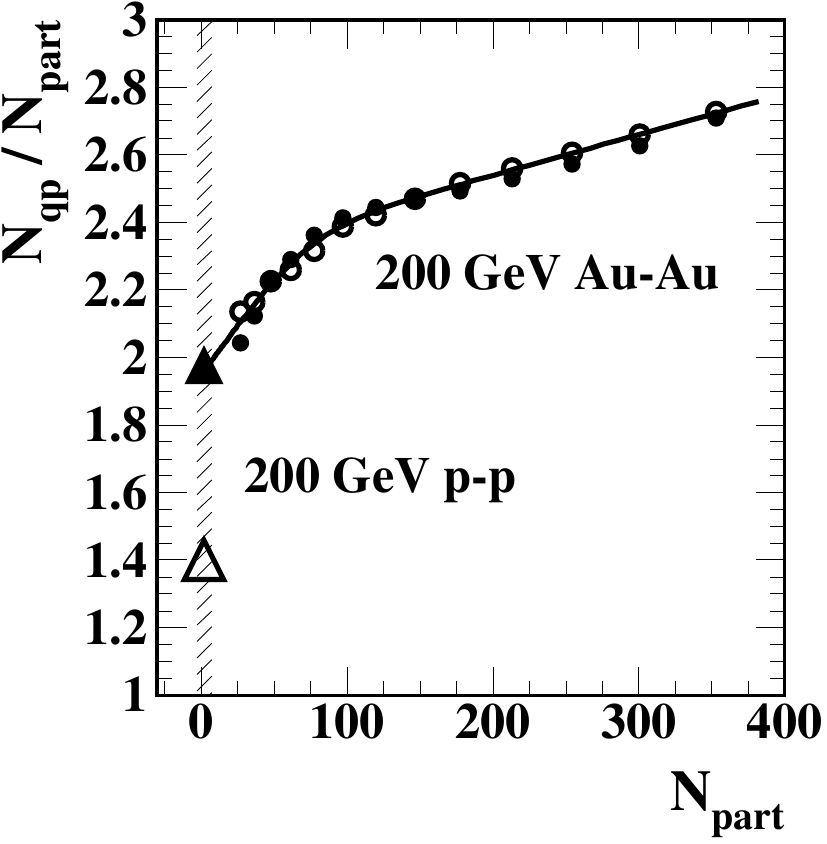}
  \includegraphics[width=1.65in]{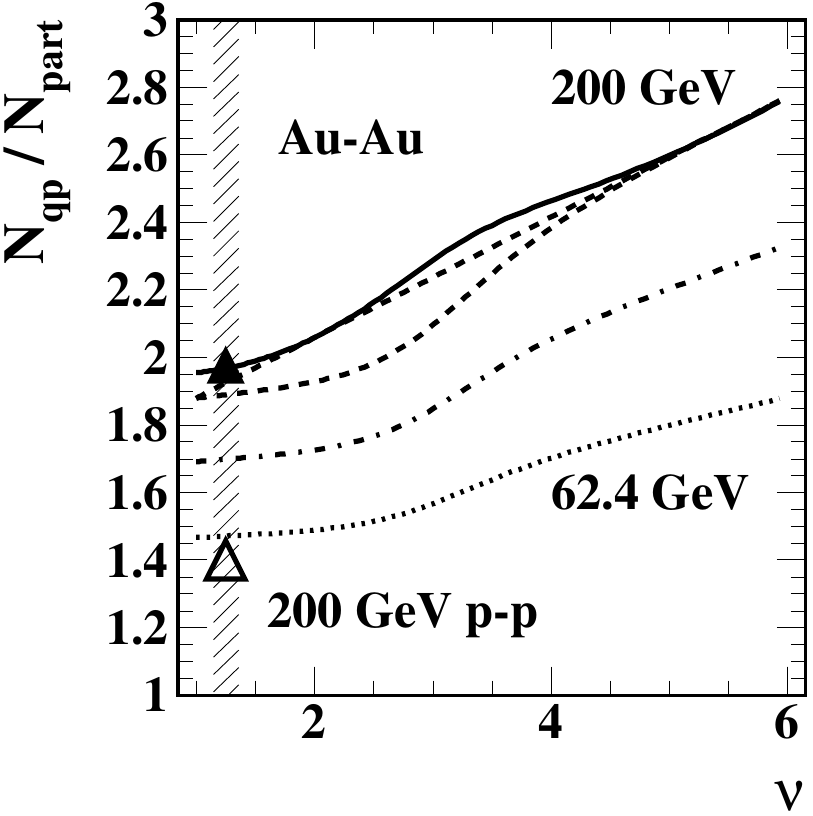}
 \caption{\label{consquark}
 Left: Ratio $N_{qp} / N_{part}$ derived from a Glauber Monte Carlo simulation of 200 GeV \auau\ collisions from Ref.~\cite{phenix}, with $N_{part}$ values from that reference (open circles) and from Eq.~(\ref{glauber}) (solid points) with a parametrization used in the present study (curve). The two values for 200 GeV \pp\ collisions (triangles) are discussed in Secs.~\ref{phminbias} and \ref{pparg}.
Right:  The ratio parametrization from the left panel vs participant pathlength $\nu$ (solid curve). Also plotted are the TCM for 200 GeV \auau\ with fixed $x = 0.1$ (upper dashed curve) and $(2/N_{part}) dn_{ch}/d\eta$ trends with variable $x$ for three energies (lower dashed curve and dash-dotted and dotted curves). Those curves are scaled down relative to data by factor 1.35.
  } 
  \end{figure}

In Ref.~\cite{phenix} the number of conjectured participant quarks $N_{qp}$ is also determined with a Glauber Monte Carlo simulation. The CQ cross section is defined so as to sum to the \nn\ cross section at each collision energy, and the number of participant CQs is then determined vs \aa\ centrality. Simulation results are shown in Fig.~\ref{consquark} (left panel) as the ratio $N_{qp} / N_{part}$ with open points ($N_{part}$ from Ref.~\cite{phenix} Table V) and solid points [$N_{part}$ from Eq.~(\ref{glauber})]. The solid curve is a simple parametrization used below for illustration. The open and solid triangles represent estimated ratios for 200 GeV \pp\ collisions. The solid point is obtained by extrapolating the solid curve. The open point is discussed in Secs.~\ref{phminbias} and \ref{pparg}. 
  
 In Sec.~V of Ref.~\cite{phenix} the $N_{qp}$ trend (solid curves) is said to be similar to that for the charged-particle yield $n_{ch}$ per participant nucleon, possibly explaining the measured $n_{ch}$ data trend without recourse to a TCM dijet contribution. In Fig.~\ref{consquark} (right panel) $(2/N_{part})dn_{ch}/d\eta$ distributions for three energies are scaled down by common factor 1.35. The upper dashed line that does approximate the $N_{qp}$ trend is the conventional TCM for 200 GeV assuming fixed parameter value $x \approx 0.1$ for all centralities. The lower dashed curve is the TCM with parameter $x$ varying from 0.015 (GLS extrapolation from \pp\ collisions) to 0.095 (derived empirically from more-central \auau\ collisions) as required by the measured spectrum and angular-correlation data described below. The lowest two curves show the same variable-$x$ TCM model derived from 130 and 62.4 GeV data. Deviations from the \pp\ GLS reference value $x \approx 0.015$ have been related quantitatively to modification of parton fragmentation to jets in more-central 200 GeV \auau\ collisions~\cite{fragevo}.

Figure~\ref{consquark} (right panel) suggests that detailed comparisons between measured hadron-production trends and the $N_{qp}$ trend for more-peripheral \aa\ collisions may already falsify the CQM, as discussed in Sec.~\ref{aaarg}. But that centrality interval is typically de-emphasized at RHIC. More-differential methods may provide a  definitive picture, as discussed in Secs.~\ref{ppcoll} and \ref{aacoll}.
       
\subsection{The two-component model} \label{2comp}

The TCM of hadron production in high-energy nuclear collisions includes (near mid-rapidity) a soft component (SC) attributed to projectile-nucleon fragmentation (dissociation) with \aa\ centrality dependence proportional to $N_{part}$ and a hard component (HC) attributed to large-angle-scattered parton fragmentation to dijets proportional to $N_{bin}$. The TCM has been applied successfully to single-particle spectra and their integrated yields~\cite{ppprd,hardspec} and to two-particle correlations~\cite{porter2,porter3,axialci,anomalous} from \pp\ and \aa\ collision systems at several energies. 

The soft component of spectra appears to be universal across a broad range of collision systems and energies, reflecting local charge conservation (unlike-sign pair correlations dominate) and an \mt\ spectrum described by a L\'evy distribution with fixed parameters~\cite{ppprd,hardspec}.
The hard component of \pp\ spectra is quantitatively consistent with predictions~\cite{fragevo} derived from measured jet (parton) energy spectra~\cite{ua1} and fragmentation functions (FFs)~\cite{eeprd}. The hard component of \pp\ angular correlations on angle differences is consistent with  theoretical expectations for dijets: (a) a same-side (SS) 2D peak at the origin corresponds to intrajet correlations and (b) an away-side (AS) 1D peak on azimuth corresponds to interjet (back-to-back jet pairs) correlations. 

A TCM reference (null hypothesis) for \aa\ collisions can be defined by extrapolation from \pp\ collision data assuming linear superposition of \nn\ collisions according to  the Glauber model, the reference then described appropriately as Glauber linear superposition~\cite{anomalous}. 
Testing the GLS reference in \aa\ collisions requires accurate measurements especially for more-peripheral collisions. Deviations from the GLS reference in more-central \aa\ collisions may then signal new physics. Monte Carlos such as HIJING (with no jet quenching~\cite{hijing}) based on models of \pp\ collisions (PYTHIA~\cite{pythia}, HERWIG~\cite{herwig}) should follow the GLS model derived from \pp\ data for all \auau\ centralities, but they do not~\cite{anomalous}. The peripheral region is typically de-emphasized in RHIC analysis, for instance by plotting data vs $N_{part}$ where the more-peripheral 50\% of the total cross section is confined to $N_{part} < 50$ or 13\% of the total axis interval. Rather than extrapolate upward from \pp\ collisions to establish a GLS reference the conventional approach at RHIC has been to derive a fixed value of TCM model parameter $x$ from more-central \auau\ data and then compare only the more-central region with \aa\ models~\cite{phobostcm,kn}.

The hard components of spectra and angular correlations from more-peripheral \auau\ collisions are indeed consistent with extrapolation of measured \pp\ structure 
but undergo substantial changes with increasing \auau\ centrality above a {\em sharp transition} (ST) on centrality near $\nu \approx 3$ or $\sigma / \sigma_0 \approx 0.5$~\cite{hardspec,fragevo,anomalous}.  Accurate spectrum and MB measurements reveal that the effective TCM $x$ from 200 GeV \auau\ increases with centrality from a smaller \pp\ value in more-peripheral collisions to six times that value in more-central collisions, the change corresponding mainly to evolution of parton FFs~\cite{fragevo}.  The strong variation of $x$ has implications for the structure of MB event-frequency or cross-section distributions on $n_{ch}$ and \et, as discussed in App.~\ref{mbstruct}.

 \subsection{Single-particle spectra in p-p and \aa\ collisions}
 
 Single-particle \pt\ or \yt\ spectra from high-energy \pp\ collisions can be grouped in multiplicity classes based on the total charge multiplicity within some angular acceptance $\Delta \eta$. Variation of the spectrum shape with multiplicity  can be employed to decompose spectra into two components with fixed shapes independent of multiplicity: ``soft'' (scaling approximately as $n_{ch}$) and ``hard'' (scaling approximately as $n_{ch}^2$)~\cite{ppprd}. 
 The soft and hard spectrum components integrated within some angular acceptance yield soft $n_s$ and hard $n_h$ multiplicity components, with $n_{ch} = n_s + n_h$. The quantitative correspondence of the \pp\ spectrum hard component to a pQCD jet description was established in Ref.~\cite{fragevo}.  

The \pp\ spectrum TCM serves in turn as the baseline or GLS reference for an \aa\ spectrum TCM in which some model parameters extrapolated from \pp\ phenomenology are permitted to vary to accommodate the \aa\ spectrum data. Whereas $n_{ch}$ is the control or ``centrality'' parameter for \pp\ collisions participant path length $\nu$ is the relevant centrality parameter for \aa\ collisions.
  
\subsection{Correlations in p-p and \aa\ collisions}

Angular correlations on $(\eta,\phi)$ and correlations on transverse-rapidity space $y_t \times y_t$ have been studied in detail for \pp\ and \aa\ collisions~\cite{axialci,axialcd,porter2,porter3,anomalous,elizabeth}. A two-component model for correlations in \pp\ collisions provides the basis for a mathematical model of angular correlations~\cite{porter2,porter3} which is then applied to \aa\ collisions with minor modifications~\cite{axialci,anomalous}. The SC produces correlations only on $\eta$, only for unlike-sign (US) charge pairs, and only below 0.5 GeV/c~\cite{porter2,porter3}. It falls to zero amplitude by mid-centrality in 200 GeV \auau\ collisions.

The hard component of angular correlations is fully consistent with expectations for dijet correlations below the sharp transition on centrality. Above the ST the same-side 2D peak (intrajet correlations) becomes elongated on $\eta$ and narrows significantly on $\phi$~\cite{axialci,anomalous}. The volume of the SS 2D peak is consistent with the SP spectrum hard component in terms of number of integrated fragment pairs, corresponding quantitatively to the pQCD-predicted number of dijets and their fragments~\cite{jetspec}. \pt\ angular correlations analogous to particle-number angular correlations are described in Sec.~\ref{ptangle}.

Particle-number correlations on transverse-rapidity space $y_t \times y_t$ from \pp\ collisions also exhibit distinct soft and hard components~\cite{porter2,porter3}. The 2D soft component lies almost completely below $p_t = 0.5$ GeV/c. The 2D hard component lies almost completely above that point and, projected onto 1D $y_t$, is consistent with the SP \yt\ spectrum hard component. The phenomenology of \yt\ correlations has been explored in detail~\cite{porter2,porter3,elizabeth}. Above the ST on \auau\ centrality the hard component splits, with one part (mainly pions) moving to smaller \yt\ and the other part (mainly protons) moving to larger \yt, again in agreement with measured SP spectrum trends~\cite{hardspec}.

 \section{PHENIX minimum-bias $\bf E_t$ data} \label{phetdata}
 
We first consider the analysis and alternative interpretation of MB distributions on \et\ presented in Ref.~\cite{phenix}.  The basic data consist of MB distributions integrated within some fiducial angular acceptance and corrected to a reference acceptance as shown in Fig.~3 and  corrected \et\ production vs centrality as  in Table V of that paper. For the present study we emphasize the 200 GeV \pp\ and \auau\ data and do not consider d-Au data or Additive Quark Model scaling. As noted, to simplify the notation we adopt the PHENIX convention $dE_t/d\eta  \leftrightarrow$ \met.

\subsection{PHENIX EMCal angular acceptance}

The absolute size of the detector angular acceptance has important consequences for MB distributions. In Ref.~\cite{phenix} a distinction is established between the fiducial (physical) detector angular acceptance and the defined reference acceptance ($\Delta \eta = 1, \Delta \phi = 2\pi$). The PHENIX EMCal has a substantially smaller angular acceptance than some comparable detectors. For instance, the effective acceptance of the PHENIX EMCal is 1/(8.4 x 1.2) $\approx$ 1/10 the STAR TPC or barrel EMCal acceptance. Certain aspects of data obtained within the fiducial acceptance such as mean values can be corrected to a reference acceptance. But the effects of acceptance on other aspects such as statistical fluctuations may persist as biases in the corrected data. Two issues are relevant for this study: (a) void probabilities and (b) structure of the {\em terminus} or central-\aa\ end (tail) of MB distributions.

In the average non-single-diffractive (NSD) \pp\ collision approximately 10 particles (charged + neutral) may fall within the STAR fiducial EMCal/TPC acceptance (with $\Delta \eta \approx 2$), but only 1 particle would fall within the PHENIX fiducial acceptance. The Poisson void probability for STAR is exp(-10) $\approx 10^{-5}$, but for PHENIX the void probability is $p_0 \approx 0.35$ (as in Table X of Ref.~\cite{phenix}). Reference~\cite{phenix} notes that ``The importance of taking account of $p_0$...can not be overemphasized.'' For the PHENIX acceptance that statement is well justified.

MB data from two detectors corrected to a common reference acceptance may be quite different in  the limit of central collisions where the tail of the distribution reflects fluctuations in central collisions determined by the actual or fiducial acceptance. The tail width relative to the terminus mean $\sigma_n / \bar n_0 \approx 1/\sqrt{\bar n_0}$ (assuming Poisson statistics) for STAR relative to PHENIX should be  $1/\sqrt{10} \approx 1/3$. Rescaling the MB distribution to the reference acceptance will not change that ratio. Thus, the terminus widths of MB distributions from PHENIX should be about three times larger than the equivalent from STAR. That issue is discussed further in App.~\ref{tailflucts}.

\subsection{$\bf \langle E_t \rangle$ centrality trends}

\begin{figure}[h]
  \includegraphics[width=1.65in,height=1.6in]{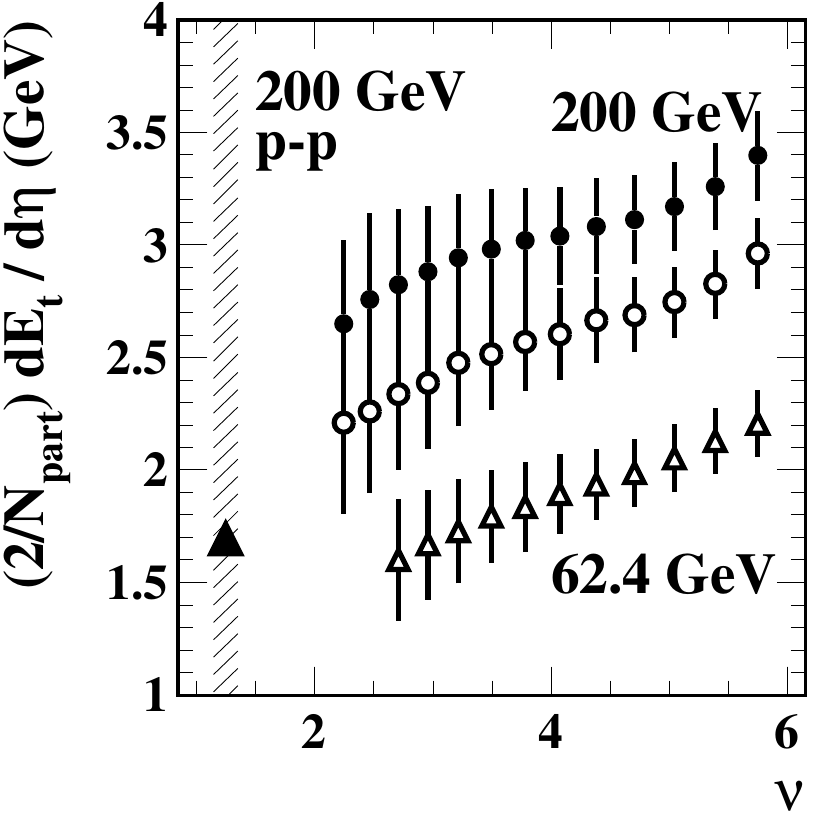}
  \includegraphics[width=1.65in,height=1.6in]{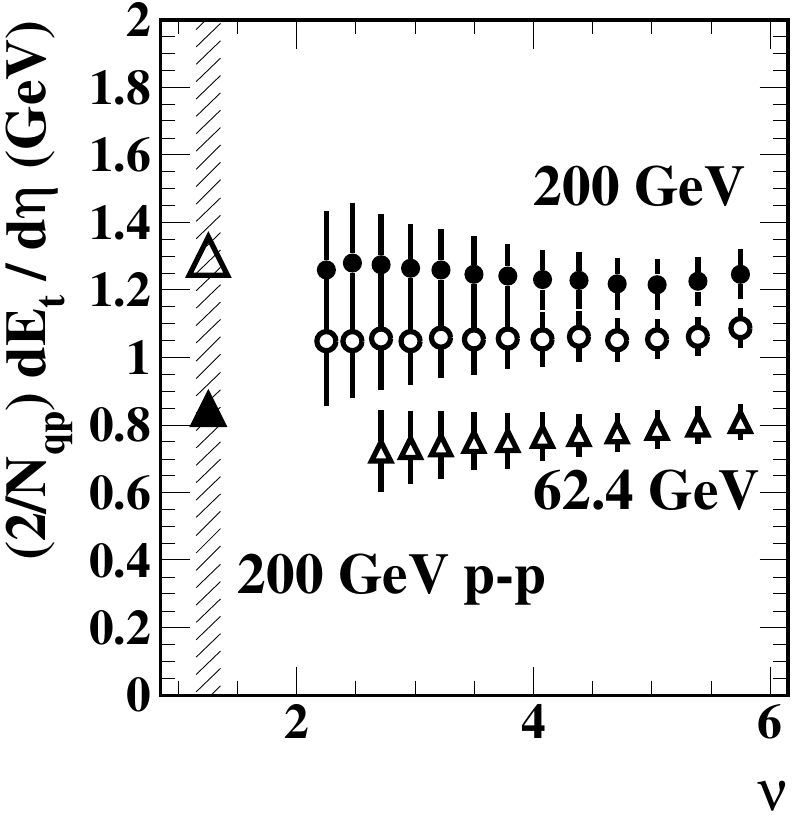}
  \caption{\label{mbet}
Left: Per-participant transverse-energy production measured by $(2/N_{part})dE_t/d\eta$ vs centrality for 200, 130 and 62.4 GeV \auau\ collisions. The data and uncertainties (stat + sys) are from Tables V-VII of Ref.~\cite{phenix}. The $N_{part}$ and $\nu$ values are from Eq.~(\ref{glauber}). The solid triangle is the corrected \pp\ value from Table X of Ref.~\cite{phenix}.
Right: Data from the left panel divided by ratio $N_{qp}/N_{part}$ from Fig.~\ref{consquark}. The \pp\ points are obtained from the point in the left panel with $N_{qp}/N_{part}$ ratio values 1.95 (solid point, extrapolation of the Fig.~\ref{consquark} solid curve) and 1.4 (open point, separate \pp\ Glauber constituent-quark simulation reported in Ref.~\cite{phenix}).
 } 
 \end{figure}

Figure~\ref{mbet} (left panel) shows ratio $(2/N_{part}) dE_t/d\eta$ from \auau\ collisions for three energies vs centrality measured by mean participant pathlength $\nu$. The $dE_t/d\eta$ data and uncertainties are from Tables V-VII of Ref.~\cite{phenix} while the $N_{part}/2$ and $\nu$ values are from Eq.~(\ref{glauber}). The uncertainties include total point-to-point plus common offset systematic uncertainties.  The \met\ centrality trends are described as varying ``nonlinearly'' with $N_{part}$, interpreted to imply inconsistency with the participant scaling expected from soft production and therefore apparently requiring hard scattering and the TCM. 

Figure~\ref{mbet} (right panel) shows ratio $(2/N_{qp}) dE_t/d\eta$ vs centrality obtained by dividing the data in the left panel by ratio $N_{qp} / N_{part}$ from Fig.~\ref{consquark}.  The ratio data above $\nu = 2$ appear to be constant within systematic uncertainties, suggesting that mid-rapidity hadron production in \auau\ collisions actually occurs by a soft process scaling with the number of  CQ participants, not with nucleon participants. Values for 200 GeV \pp\ collisions derived from data in Table X of Ref.~\cite{phenix} are plotted at $\nu = 1.25$ ($\approx$ NSD \nn\ collisions). The upper \pp\ datum is said to confirm the self-consistency of the CQ analysis. The two \pp\ results are related to \pp\ $N_{qp}$ values in Fig.~\ref{consquark} and are discussed in Secs.~\ref{phminbias} and \ref{pparg}.  

\begin{figure}[h]
  \includegraphics[width=1.65in]{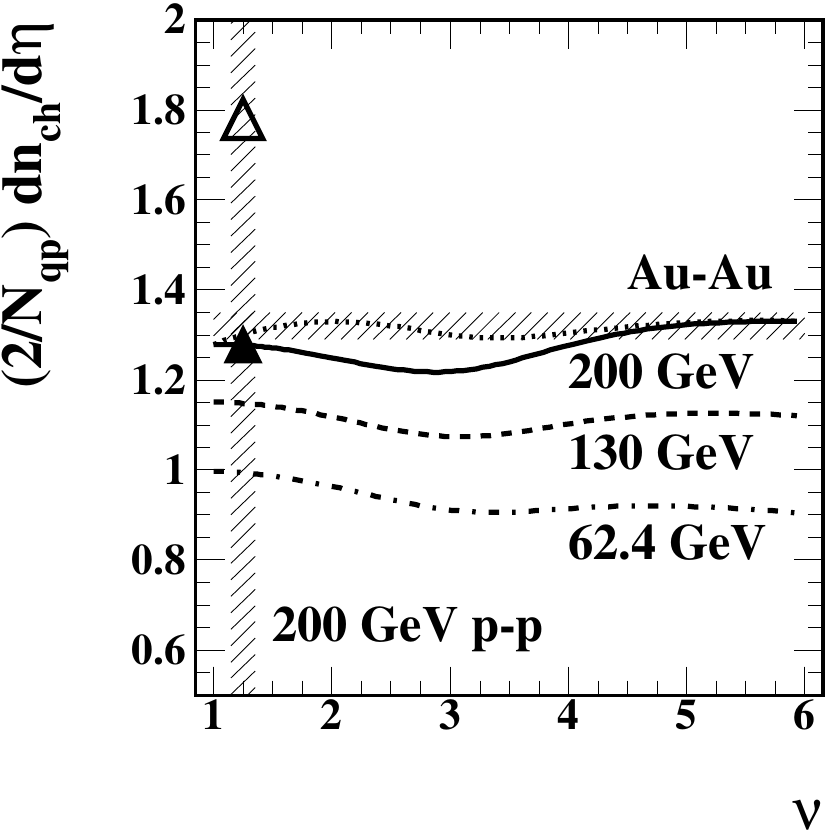} 
   \includegraphics[width=1.65in]{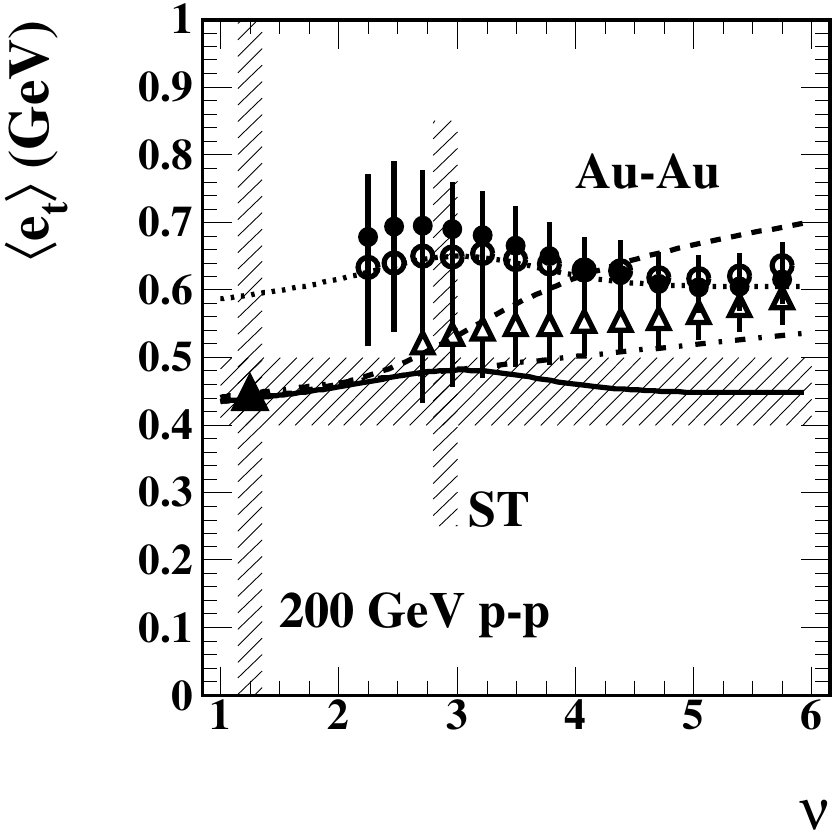}
\caption{\label{met}
Left: The relation of measured charged-hadron production to Glauber-simulated number of constituent quarks in \auau\ collisions for three energies derived from the curves in Fig.~\ref{consquark}.  The points are for 200 GeV \pp\ collisions.
Right: Per-hadron mean $E_t$ defined as $\langle e_t \rangle = dE_t/d\eta \, /\, dn_{tot}/d\eta$, with $n_{tot}$ (all hadrons contributing to $E_t$) approximated by $1.5 n_{ch}$ assuming that pions dominate (approximately 85\% of all hadrons). The $dE_t/d\eta$ data and uncertainties are taken from Tables V-VII of Ref.~\cite{phenix}. The $dn_{tot}/d\eta$ trends are derived from STAR data represented by TCM parametrizations with varying $x$ in Fig.~\ref{consquark} (right panel). The solid curve is a prediction from the TCM for 200 GeV \auau\ collisions~\cite{fragevo}. Other curves are described in the text.
 } 
 \end{figure}

Figure~\ref{met} (left panel) shows the TCM for $dn_{ch}/ d \eta$ data from three collision energies similarly scaled by $N_{qp}$. The dotted curve represents the conventional TCM with fixed $x \approx 0.1$, while the other curves represent variable-$x$ models that describe spectrum yields accurately. The dotted curve appears to be constant within typical systematic uncertainties, seeming to confirm a claim by Ref.~~\cite{phobostcm} that the increase in  $(2/N_{part}) dn_{ch}/d\eta$ described by the fixed-$x$ TCM arises not from increasing contributions by hard processes but rather from increasing CQ number (and associated soft hadrons) relative to participant nucleons. However  the lowest three curves, representing measured production trends, consistent below  the ST with the GLS reference extrapolated from \pp\ measurements, deviate substantially from the CQM for more-peripheral collisions and may falsify that model (see Sec.~\ref{aaarg}). The two points are \pp\ values derived from the two \pp\ values of $N_{qp}$ from Fig.~\ref{consquark}. The preferred value $N_{qp} = 2.8$ from a \pp\ Glauber simulation leads to the open point which is far above the trend for \auau\ collisions.

As a logical next step we bypass the conjectured geometry model parameters by calculating \meet\  $= (dE_t/d\eta)/( dn_{tot}/d\eta)$, where $n_{tot}$ represents all hadrons contributing to \met,  approximated in this case by $n_{tot} \approx 1.5\, n_{ch}$ to account for $\pi^0$s.
Figure~\ref{met} (right panel) shows \meet\  vs \auau\ centrality (points with uncertainties) for three collision energies. The $n_{tot}$ values are derived from the full TCM with variable $x$ that describes accurately the measured charged-particle spectrum integrals~\cite{hardspec}. Also shown is the \meet\ value for 200 GeV \pp\ collisions (solid triangle) derived from the PHENIX corrected \met\ value for \pp\ collisions  from Table X ($2.6 \times 0.65 = 1.7$ GeV). That value is consistent with NSD \pp\ charge-particle \mpt\ $\approx 0.4$ GeV/c given $dn_{ch}/d\eta \approx 2.5$ with additional factor 1.5 to approximate total hadrons. 

If free-streaming hadrons emerge from a thermalized medium, \meet\ or \mpt\ is expected to relate to a temperature T that may vary with \aa\ centrality. However, the concept of an equilibrated temperature for \auau\ collisions below the sharp transition that appear to exhibit negligible particle rescattering is questionable.

The \meet\ centrality trend can be predicted from the TCM based on measured spectrum properties~\cite{hardspec} as
\bea \label{meteq}
\langle e_t \rangle(\nu) &\approx& {[1-x(\nu)] \langle e_t \rangle_{soft} + x(\nu)\, \nu\,  \langle e_t \rangle_{hard}(\nu)},
\eea
where $x(\nu)$ is the same centrality-dependent parameter that describes integrated  yields $n_{ch}(\nu)$, $ \langle e_t \rangle_{soft}$ is the fixed mean value for soft hadron production and $ \langle e_t \rangle_{hard}(\nu)$ is the mean value for the hard component that varies strongly with centrality due to modification of parton fragmentation. Below the \auau\ ST $x$ and $ \langle e_t \rangle_{hard}$ have fixed values derived from \pp\ collision data. 

To facilitate comparisons we approximate all hadrons as pions. The effect of the pion mass is small. \meet\ is only a few percent larger than \mpt\ for \meet\ $\approx 0.5$ GeV so we increase  measured \mpt\ values by 0.05 GeV/c accordingly. The $\langle p_t \rangle_{soft}$ value for NSD \pp\ collisions (see Fig.~\ref{ppcomm} -- right panel) is 0.385$\pm 0.02$ GeV/c, and the $ \langle p_t \rangle_{hard}$ value is 1.2$\pm 0.1$ GeV/c. In Fig.~\ref{met} the horizontal hatched band estimates $\langle e_t \rangle_{soft}$ and the dash-dotted curve represents the GLS reference \meet\ prediction. We observe that the hadron hard-component multiplicity increases much faster than GLS above the sharp transition (vertical band ST). If  $ \langle e_t \rangle_{hard}$ remained fixed while $x(\nu)$ increased according to $dn_{ch}/d\eta$ measurements the predicted \meet\ trend would be the dashed curve. However, spectrum analysis reveals that  \mpt$_{hard}$ falls from 1.2 GeV/c below the ST to 0.6 GeV/c for more-central \auau\ collisions. If that variation is also incorporated the TCM prediction for \meet\ becomes the solid curve.

The PHENIX 200 and 130 GeV \auau\ data trends are consistently 35\% above the solid curve, whereas the \meet\ value for \pp\ collisions (solid triangle) is consistent with STAR \mpt\ data (hatched band). It is notable that the point-to-point variation of PHENIX \meet\ data follows the TCM prediction closely. The dotted curve is the TCM solid curve multiplied by factor 1.35.  The TCM was not otherwise adjusted to accommodate those data.

\subsection{Minimum-bias distributions on $\bf E_t$} \label{phminbias}

Given those results for the \auau\ centrality dependence of \et\ mean values, Ref.~\cite{phenix} extends the CQ treatment to full MB distributions on \et. Several hypotheses are considered in which $E_t$ production scales with some  element number $N_X$, where $X$ can represent \nn\ binary collisions, participant (wounded) nucleons, constituent quarks or participant-quark-related strings. A model MB distribution on an observable such as \et\ is generated by folding a MB distribution on $N_x$ with model conditional MB distributions on \et\ given $N_x$ elements. The result is described as the Extreme Independent Model. 

The \aa\ MB distribution on $N_X$ is determined by Glauber Monte Carlo simulations. The measured MB distribution on $E_t$ for \pp\ collisions is used to infer an equivalent for element type $X$ denoted $f_{1X}(E_t)$. Conditional distribution $P_{N_X}(E_t)$ for $N_X$ elements is obtained by convoluting $f_{1X}(E_t)$ multiple times and combining with calculated weights. Finally, the \aa\ MB distribution on \et\ is obtained by convoluting the MB distribution on $N_X$ with the  $P_{N_X}(E_t)$ according to Eq.~(7) of Ref.~\cite{phenix}.

The model chosen for $f_{1X}(E_t)$ is the gamma distribution, for which a simple parameter change conveniently describes the results of multiple convolutions. Aside from the algebraic convenience the gamma distribution on \et\ does not describe the \pp\ MB distribution well; the tail is poorly represented. The importance of accounting for void events (\nn\ events with no $E_t$ in the fiducial acceptance, described by probability $p_0$) is emphasized. 

The \auau\ Glauber MB distributions on $N_{part}$ and $N_{bin}$ are accurately represented by Eq.~(\ref{glauber}). The MB distribution on $N_{qp}$ is shown in Fig.~9 of Ref.~\cite{phenix}. Figure~10 (b) of Ref.~\cite{phenix} demonstrates an unfolding of the \pp\ MB distribution to obtain $f_{1qp}(E_t)$. The decomposition is based on $w_n$ numbers in Table XIII equivalent to $N_{qp} = 2.8$ for \pp\ collisions. That value can be contrasted with $N_{qp} = 4$ obtained by extrapolating the $N_{qp} / N_{part}$ trend in Fig.~\ref{consquark} to $\nu = 1.25$ (\nn\ collisions). The difference is discussed in Sec.~\ref{pparg}. The MB endpoint on $N_{part}$ is 382, whereas that on $N_{qp}$ is approximately $382 \times 2.8 = 1070$. The CQM strongly disagrees with the tail of the \pp\ data distribution as noted. Those results are combined per Eq.~(7) or (12) to obtain the model MB distribution compared with \auau\ data in Fig.~11. The model is said to describe the MB data well. The ``excellent agreement'' is interpreted to support the CQM.

\subsection{The TCM as expressed within a CQM context} \label{tcmcq}
 
 In the CQM context a significant role for dijet production in the MR is considered unlikely based on historical experience:  ``This [hard scattering in the TCM] seems to contradict the extensive measurements of $N_{ch}$ and \et\ distributions in p+p collisions described in Sec. II [of Ref.~\cite{phenix}] which show that these distributions represent measurements of the `soft' multiparticle physics that dominates the p+p inelastic cross section''~\cite{phenix}.  Reference~\cite{phenix} proposes relations between the TCM and (a) production centrality trends and (b) MB distributions on \et\ based on its Eqs.~(6,21)  interpreted to demonstrate the unsuitability of the TCM for description of data.  
 
 The conjectured TCM for \et\ production includes  a single-component factor \met$^{pp}$ to represent \pp\ collisions
\bea \label{phenixet}
\langle E_t \rangle^{AA} &=& \langle E_t\rangle^{pp} [(1-x) N_{part}/2 + x N_{bin} ],
\eea 
implying that $E_t$ production for \pp\ soft and hard components is the same (a common factor).   If TCM parameter $x$ in Eq.~(\ref{phenixet}) also describes hadron yields then the equation reduces to \meet$^{AA}$ = \meet$^{pp}$ and $E_t$ production per hadron is assumed to be independent of \aa\ centrality. But those assumptions are contrary to the TCM with variable $x$ and contradicted by differential spectrum and correlation data. Equation~(\ref{phenixet}) can be contrasted with Eq.~(\ref{meteq}) (and see App.~\ref{etdata}) where soft and hard \meet\ components in \pp\ collisions are quite different, and the \meet\ hard component is allowed to vary with \auau\ centrality consistent with measured spectrum HC evolution~\cite{hardspec}.

The relation of the TCM to MB distributions is considered to be especially problematic. Analogies are drawn from conjectured TCM descriptions of production centrality trends. It is first assumed that the data MB distribution is the weighted sum of the limiting cases (MB on $N_{part}$ and $N_{bin}$) by analogy with the TCM for integrated yields [e.g.\ Eq.~(\ref{phenixet}) above].  One example  in Figs.~17 and 18 of Rev.~\cite{phenix} is represented by the first line of
 \bea \label{phenixmbeq}
 \frac{d\sigma}{dE_t} &=& (1-x)\left(\frac{dE_t}{dN_{part}}\right)^{-1}\frac{d\sigma}{dN_{part}}
 \\ \nonumber
 &+& x\left(\frac{dE_t}{dN_{bin}}\right)^{-1} \frac{d\sigma}{dN_{bin}},
 \\ \nonumber \text{or}~
 &=&\left[\frac{(1-x)\, dE_t}{ dN_{part}}\right]^{-1}\frac{d\sigma}{dN_{part}} 
  \\ \nonumber
  &+& \left(\frac{x\, dE_t}{dN_{bin}}\right)^{-1} \frac{d\sigma}{dN_{bin}},
 \eea 
with the same value $dE_t/dN_x \rightarrow \langle E_t \rangle^{pp} \approx 1.8$ GeV for $N_{part}$ and $N_{bin}$.  Figure~19 of Ref.~\cite{phenix} shows an alternative hypothesis, scaling the limiting cases horizontally on \met\ by fractions $x$ and $1-x$ as in the second line.  Since neither expression corresponds to the MB data the TCM is rejected. But  the apparent disagreements between  CQM {\em conjectured} TCM implementations and MB data come not from inherent failings of the TCM  but from incorrect implementations, as discussed further in Sec.~\ref{cqmtcm}.
    
The TCM does appear to describe production vs centrality trends accurately, as demonstrated in Sec.~\ref{speccorr}.  But that model is said to be ``just a proxy for the correct description of the underlying physics, because $dE_T^{AA}/d\eta$ is strictly proportional to $N_{qp}$....'' The TCM is described as an ``empirical proxy'' referring to the  conventional RHIC TCM implementation with fixed $x \approx 0.1$ fitted to more-central \auau\ data~\cite{kn}. But the TCM for production centrality dependence is not inferred empirically from a relation like Eq.~(6) of Ref.~\cite{phenix}. It is predicted from measured \pp\ TCM trends and the Glauber model that serves as a null hypothesis (GLS) for more-central \aa\ collisions. The proper TCM for MB distributions describes MB data accurately as demonstrated in App.~\ref{etdata}, whereas the CQM cannot produce a GLS equivalent.
  
 We now examine comprehensive yield, spectrum and correlation systematics from RHIC \pp\ and \auau\ collisions in the context of the TCM. We demonstrate that the TCM provides a remarkably accurate description of an extensive data phenomenology. Those data include dijet production trends that are quantitatively consistent with pQCD theory and thus inconsistent with the CQM.

 \section{\ppsec\ collisions and minijets} \label{ppcoll}
 
The TCM derived from \pp\ spectrum data and extrapolated to \aa\ centrality provides a reference for \aa\ collisions representing Glauber linear superposition. The hard-component contribution to the \pp\ TCM is attributed to MB dijets. Because of the steepness of the QCD parton spectrum most dijets appear near  the most-probable jet energy 3 GeV (defining minijets)~\cite{ua1}. Issues relating to \pp\ collision centrality, hadron and dijet production and the TCM are discussed in~\cite{pptheory}.

\subsection{Soft and hard events and yield $\bf n_{ch}$ components} \label{tcmyield}

\pp\ collisions can be classified as soft or hard event types. Soft and hard events are distinguished from soft and hard components of ensemble-averaged yields, spectra and correlations. Soft events include no jet structure within the acceptance and therefore only a soft component. Hard events include at least one minimum-bias dijet within the angular acceptance and therefore both soft and hard spectrum and correlation components.  

We observe that dijet production in \pp\ collisions scales approximately as $n_{ch}^2$~\cite{ppprd} and is most directly related to the multiplicity soft component $n_s$. For a given $n_s$ the dijet number within some $\eta$ acceptance $\Delta \eta$ is $n_j(n_s) = \Delta \eta\, f(n_s)$, with dijet frequency $f(n_s)$ per unit $\eta$ scaled from non-single-diffractive (NSD) \pp\ collisions. The relation between soft and hard components $n_s$ and $n_h$ and dijet rate $n_j(n_s)$ is established in Sec.~\ref{dijet}.  The Poisson probabilities for soft and hard events are then respectively $P_s(n_s) = \exp(-n_j)$ and $P_h(n_s) = 1- P_s(n_s)$. For small $n_j$ $P_h \approx n_j$. For the PHENIX acceptance the probability of a dijet in NSD \pp\ collisions is $P_h \approx 0.004$.

The yields $n_x$ defined here correspond to spectrum integrals within some angular acceptance $2\pi$ and $\Delta \eta$. For each multiplicity class 
defined in terms of an $n_{ch}$ interval we have $n_s + n_h = n_{ch}$ averaged over all events. For soft events $n_s'' = n_{ch}$ and for hard events $n_s' + n_h' = n_{ch}$. We then obtain the following relations:
\bea \label{tcmpp}
n_{ch} &=& n_s + n_h = P_s n_{ch} + P_h (n_s' + n_h') \\ \nonumber
n_s &=& P_s n_{ch} + P_h n_s' ~~\text{and}~~ n_h = P_h n_h' 
\eea

\subsection{p-p single-particle spectra} \label{ptspec} \label{ppspecc}

Single-particle spectra from 200 GeV \pp\ collisions plotted on $y_t$ for several $n_{ch}$ classes reveal a composite spectrum structure represented by two fixed functional forms  [unit-integral soft and hard components $\hat S_0(y_t)$ and $\hat H_0(y_t)$] with amplitudes scaling approximately as $n_{ch}$ and $n_{ch}^2$~\cite{ppprd}. The combination defines the TCM for $y_t$ spectra from \pp\ collisions conditional on measured $ n_{ch}$ integrated within some $\eta$ acceptance $\Delta \eta$ described by
\bea \label{ppspec}
\rho(y_t,n_{ch}) \equiv \frac{dn_{ch}}{y_t dy_t \Delta \eta} \hspace{-.05in}&=&\hspace{-.05in} S(y_t,n_{ch}) + H(y_t,n_{ch}) \\ \nonumber
&=&  \rho_s( n_{ch}) \hat S_0(y_t)  +  \rho_{h}( n_{ch}) \hat H_0(y_t),
\eea
where $\rho_s = n_s / \Delta \eta$ and $\rho_h = n_h / \Delta \eta$ are soft and hard angular densities, and $\rho_0(n_{ch}) = n_{ch} / \Delta \eta$ is the corresponding  total charge density. Soft component $\hat S_0(y_t)$ is defined as the limiting form as $\rho_s \rightarrow 0$ of spectra normalized as $\rho / \rho_s$. Hard component $\hat H_0(y_t)$ models data hard components $H(y_{t},n_{ch})/\rho_s$ obtained by subtracting soft-component model $\hat S_0(y_t)$ from those normalized spectra. 

The hard component of 1D SP spectra, interpreted as a manifestation of MB dijet structure, is quantitatively consistent with jet-related two-particle correlations and pQCD predictions~\cite{ppprd,hardspec,fragevo,jetspec}.  The fixed hard-component spectrum shape $\hat H_0(y_t)$ is predicted by measured \ee\ fragmentation functions convoluted with a minimum-bias pQCD parton (dijet) spectrum with lower bound near 3 GeV and spectrum integral $\sigma_{dijet} = 4\pm1$ mb~\cite{fragevo}. 

\subsection{p-p minimum-bias dijet production} \label{dijet}

Equation~(\ref{ppspec}) integrated over some angular acceptance $\Delta \eta$ becomes
\bea \label{ppspec2}
\frac{d n_{ch}}{y_t dy_t} &=& n_s \hat S_0(y_t) + n_h \hat H_0(y_t).
\eea
The multiplicity trend for the extracted hard components reported in Ref.~\cite{ppprd} implies that $n_h/n_s = \alpha\, n_s$ with $\alpha \approx 0.005$ within $\Delta \eta = 1$ or 
\bea \label{freq}
\rho_h(n_{s}) &=& 0.005 \rho_s^2 \\ \nonumber
&\equiv& f(n_{s}) \epsilon(\Delta \eta) 2 \bar n_{ch,j},
\eea
where the second line represents the jet hypothesis from Ref.~\cite{ppprd} and defines dijet frequency $f = dn_j/ d\eta$ (dijet number per \pp\ event per unit $\eta$) with mean dijet fragment multiplicity $2\bar n_{ch,j}$ into $4\pi$ acceptance. Factor $\epsilon(\Delta \eta) \in [0.5,1]$ represents the average fraction of a dijet that appears in hard events within acceptance $\Delta \eta$~\cite{fragevo}.

Scaling from the trend in Eq.~(\ref{freq}) we can estimate $f(n_s)$ for NSD \pp\ collisions
\bea \label{f}
f_{NSD} &=& \frac{0.005}{\epsilon(\Delta \eta) 2\bar n_{ch,j}} \rho_{s,NSD}^2 \approx 0.02\pm0.005
\eea
assuming $\rho_{s,NSD}  \approx 2.5$ for 200 GeV NSD \pp\ collisions and $\epsilon(\Delta \eta)\, 2\bar n_{ch,j} \approx 1.5$ from \ppbar\ measurements~\cite{eeprd,ppprd}.  That result can be compared with a pQCD prediction for \pp\ NSD collisions $f_{NSD} = \sigma_{dijet} / \sigma_{NSD} \times \Delta \eta_{4\pi} \approx 4 / 36.5 \times 5 = 0.022 $~\cite{fragevo}. 
Thus, from a TCM analysis of the $n_{ch}$ dependence of \pp\  \yt\ spectra we obtain the model functions $\hat S_0(y_t)$ and $\hat H_0(y_t)$, the soft and hard hadron densities $\rho_s$ and $\rho_h \approx (0.006\pm0.001) \rho_s^2$ and  the dijet $\eta$ density per \pp\ event $f(n_{s}) = n_j(n_{s})/\Delta\eta \approx 0.0035 \rho_s^2$.

\subsection{p-p mid-rapidity production trends} \label{ppmb}

\begin{figure}[h]
  \includegraphics[width=1.65in]{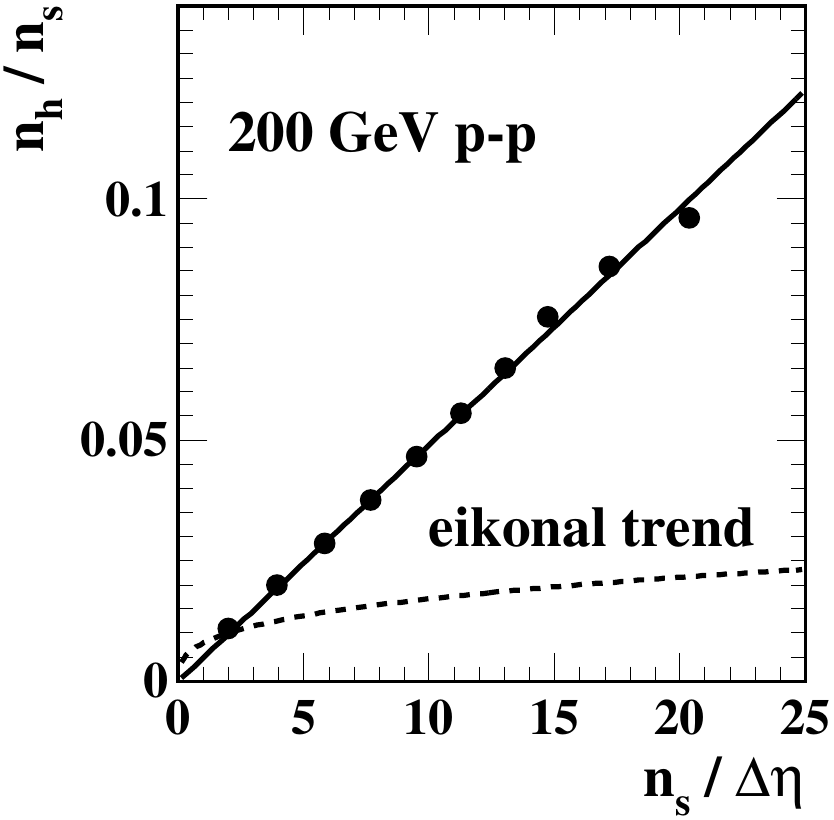}
  \includegraphics[width=1.65in]{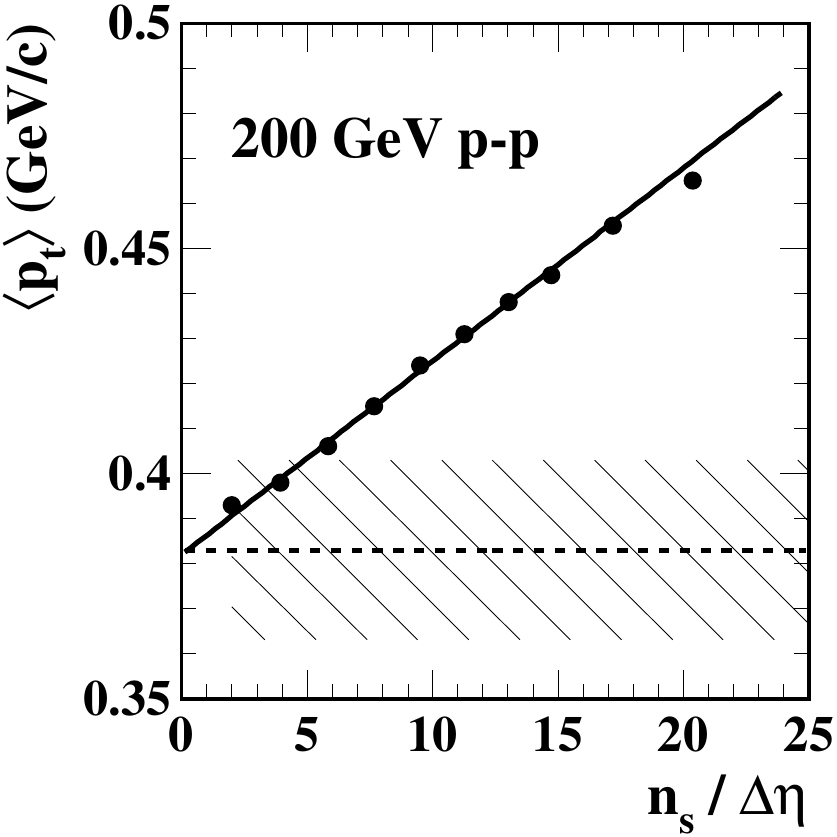}
\caption{\label{ppcomm} 
Left: Hard/soft multiplicity ratio $n_h / n_s$ (points) vs soft component $n_s$ consistent with a linear trend (line). Those data are the basis for Eq.~(\ref{freq}). Assuming that $n_s$ represents the density of small-$x$ participant partons (gluons) and $n_h$ represents dijet production by parton scattering, an eikonal model of \pp\ collision geometry (analogous to the Glauber model of \aa\ collisions) would predict the $n_s^{1/3}$ trend (dashed curve).
Right: The \mpt\ trend predicted by the \pp\ spectrum TCM (line) and as determined by direct spectrum integration (points). The hatched band represents the uncertainty in extrapolating  spectra to zero momentum.
 }  
 \end{figure}

Figure~\ref{ppcomm} (left panel) shows the data trend that lead to Eq.~(\ref{freq}). The points are determined by running integrals of spectrum data to obtain integrated multiplicities $n_s$ and $n_h$ with minimal model bias. The ratios do not depend on a spectrum model except for the soft-component L\'evy distribution used to extrapolate the \yt\ spectra to zero momentum. The fixed soft model is common to all multiplicity classes. The solid  line is $n_h / n_s = 0.005 n_s / \Delta \eta$ for acceptance $\Delta \eta = 1$. The coefficient changes to 0.006 for the STAR TPC acceptance with $\Delta \eta = 2$.
Given an accurate TCM reference for \pp\ SP spectra we can extract an analytic \mpt\ trend that may provide a baseline for \meet\ data from \auau\ collisions.

Figure~\ref{ppcomm} (right panel) shows \mpt\ values also extracted as limits of running integrals of SP spectra. The solid line is the TCM weighted-mean expression [see Eq.~(\ref{meteq})] $\langle p_t \rangle = [ (1 - x) \times0.38 + x \times 1.2$] GeV/c with $x = n_h / n_s \approx  0.005 n_s/\Delta \eta$, where the weights are approximately $n_s / n_{ch}$ and $n_h / n_{ch}$, 0.385 GeV/c is an estimate of  soft component \mpt$_{soft}$ and 1.2 GeV/c estimates  hard component \mpt$_{hard}$, slightly above the hard-component peak mode at 1 GeV/c due to its skewness (QCD power-law tail). The uncertainty in the soft-component \mpt\ (hatched band) is dominated by uncertainty in the spectrum extrapolation to zero momentum, since 30\% of the spectrum integral lies below 0.2 GeV/c.

In conventional descriptions of scattered-parton fragmentation to collimated jets and projectile-hadron fragmentation (dissociation) to hadrons along the collision axis there is a direct correspondence between partons at the bottom of a fragmentation cascade and the final-state hadron distribution denoted by Local Parton-Hadron Duality (LPHD)~\cite{lphd}. In the LPHD description the small-$x$ partons described by a parton distribution function or PDF should correspond quantitatively to soft hadron production at mid-rapidity, and dijet production at mid-rapidity should correspond to counterpropagating fluxes of small-$x$ partons within the colliding hadron projectiles.
 
The detailed correspondence between soft and hard spectrum components represented by Fig.~\ref{ppcomm} suggests that (a) small-$x$ gluons represented by $n_s$ provide the underlying degree of freedom in high-energy \pp\ collisions and (b) the dijet production trend represented by $n_h \propto n_s^2$ falsifies the eikonal approximation as applied to \pp\ collisions.  According to the Glauber model applied to \aa\ collisions and based on the eikonal approximation the expectation for \pp\ collisions should be $n_h/n_s \propto n_s^{1/3}$ (dashed curve in Fig.~\ref{ppcomm} -- left panel). The data trend implies that each participant parton in one projectile can interact with any participant in the other projectile~\cite{pptheory}.

 \subsection{p-p  two-particle correlations}
 
 Two-particle correlations for 200 GeV \pp\ collisions are fully consistent with the SP \yt\ spectrum results described above and follow the same TCM quantitatively The correlation measure $\Delta \rho / \sqrt{\rho_{ref}}$  is a 2D density proportional to the number of correlated pairs per final-state hadron~\cite{anomalous} and is analogous to ratio $n_h / n_s$ assuming $n_h \rightarrow$ correlated-pair number. 

\begin{figure}[h]
  \includegraphics[width=1.65in,height=1.4in]{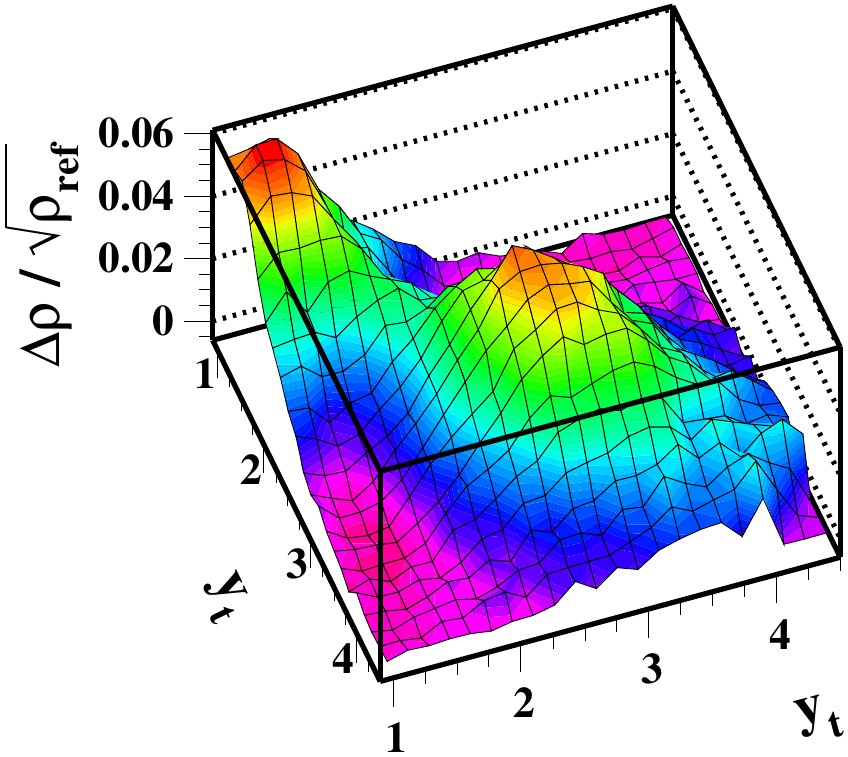}
  \includegraphics[width=1.65in,height=1.4in]{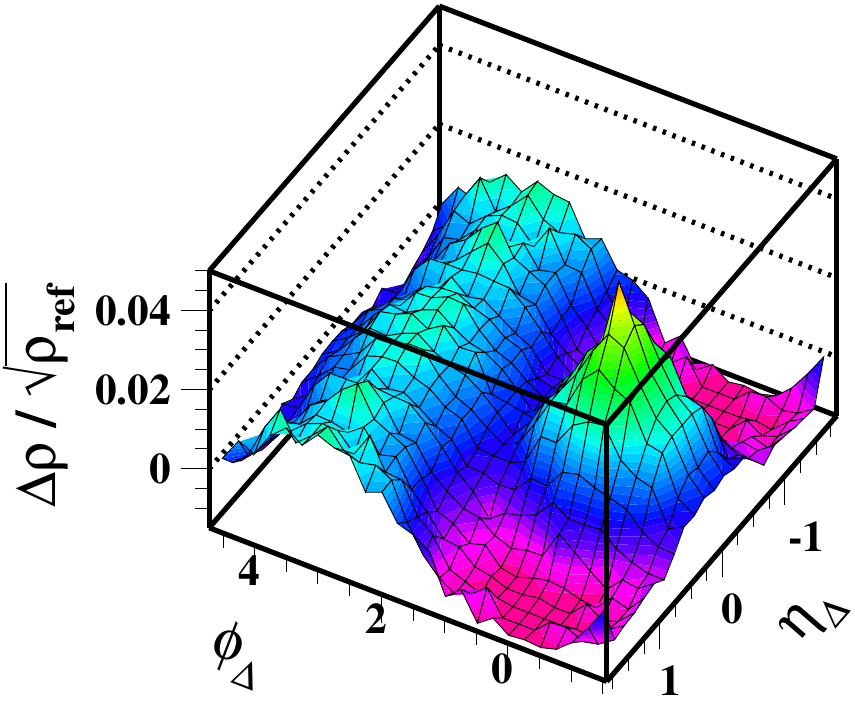}
\caption{\label{ppcorr} (Color online)
Left: Minimum-bias correlated-pair density on 2D transverse-rapidity space $y_t \times y_t$ from 200 GeV \pp\ collisions showing soft (smaller \yt) and hard (larger \yt) components as peak structures.
Right:  Correlated-pair density on 2D angular difference space $(\eta_\Delta,\phi_\Delta)$. Hadrons are selected with $p_t \approx 0.6$ GeV/c ($y_t \approx 2$). Nevertheless, features expected from dijets are observed: (a) same-side 2D peak representing intrajet correlations and (b) away-side 1D peak on azimuth representing interjet (back-to-back jet) correlations.
 }  
 \end{figure}

Figure~\ref{ppcorr} (left panel) shows $y_t \times y_t$ correlations for 200 GeV NSD \pp\ collisions. The logarithmic interval $y_t \in [1,4.5]$ corresponds to $p_t \in [0.15,6]$ GeV/c. The two peak features correspond to TCM soft and hard components. The 2D hard component with mode near \yt\ = 2.7 (1 GeV/c) corresponds quantitatively to the 1D SP spectrum hard component modeled by $\hat H_0(y_t)$. The soft component is consistent with longitudinal fragmentation (dissociation) of projectile nucleons.

Figure~\ref{ppcorr} (right panel) shows 2D angular correlations on difference variables (e.g.\ $\eta_\Delta = \eta_1 - \eta_2$). The hadron \pt\ values for that plot are constrained to lie near 0.6 GeV/c (just above \yt\ = 2), corresponding to the saddle between soft and hard components in the left panel. Although the hadron \pt\ is very low the structures expected for jet angular correlations are still clearly evident: a SS 2D peak at the origin representing intrajet correlations and a 1D peak on azimuth corresponding to interjet correlations between jets from parton collision partners. The soft component, a narrow 1D Gaussian on $\eta_\Delta$ including only unlike-sign charge pairs, is excluded by the $p_t > 0.5$ GeV/c cut. There are no ``long-range'' correlations on $\eta$ corresponding to the soft component.

 \section{\aa\ hard-component Systematics} \label{aacoll}
 
 Given the detailed reference TCM from \pp\ collisions we now consider TCM evolution with \aa\ centrality and some implications for hadron and \et\ production.  The inferred \aa\  spectrum soft component is consistent with that derived from  \pp\ spectra and appears to represent a universal transverse property of fragmentation.  
    
  \subsection{\aa\ integrated yields}
    
 The TCM for \pp\ collisions is controlled by the SC multiplicity $n_s$, conjectured to represent the number of participant small-$x$ gluons.  As noted, HC production proportional to number of participant-parton binary collisions scales as $n_s^2$. The ratio of binary collisions to participants is then proportional to the number of participants, $n_h / n_s \propto n_s$~\cite{pptheory}. 
 
 By analogy the TCM for \aa\ collisions is controlled by the SC scaling with participant multiplicity $N_{part}$.
  The \aa\ hard component HC should scale with the number of \nn\ binary collisions $N_{bin}$ estimated in the Glauber model (based on the eikonal approximation) to vary as $N_{bin} \propto N_{part}^{4/3}$.  For yields and spectra scaled by number of participant pairs $N_{part}/2$ the HC should vary proportional to $\nu \equiv 2 N_{bin} / N_{part} \approx N_{part}^{1/3}$, consistent with the eikonal approximation (whereas \pp\ collisions do not follow the eikonal approximation). The TCM for \aa\ yields within some acceptance $\Delta \eta$ is
   \bea \label{tcmaa}
n_{ch}(\nu) &=&n_s \frac{N_{part}}{2} +  n_h(\nu) N_{bin}
 \eea
 with the limiting case of Eq.~(\ref{tcmpp}). For the GLS reference $n_h$ retains the fixed value from NSD \pp\ collisions. We now focus on HCs for \aa\ spectra and correlations.

\subsection{Au-Au single-particle spectra} \label{aaspectra}

\begin{figure}[h]
  \includegraphics[width=1.65in]{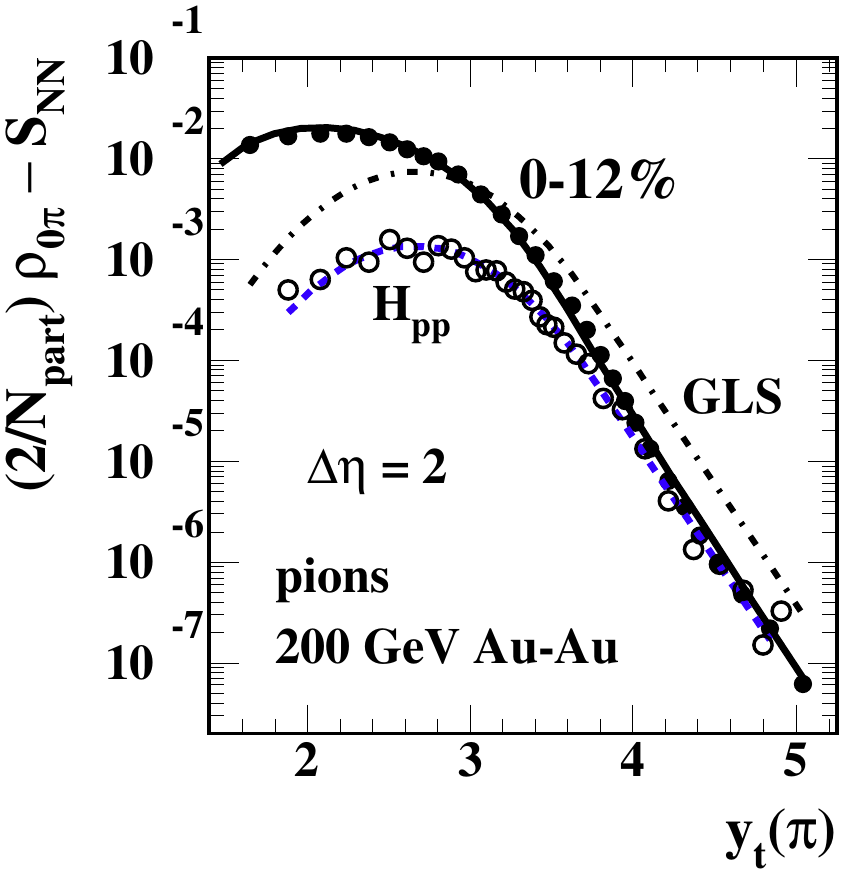}
  \includegraphics[width=1.65in]{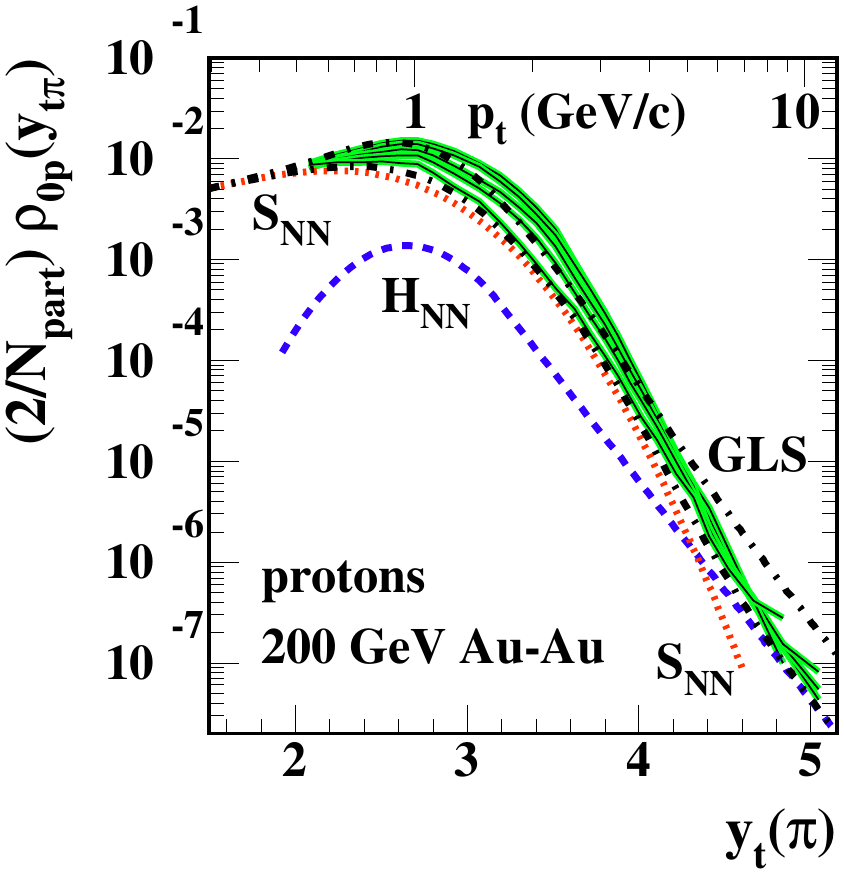}
\caption{\label{aaspec} (Color online)
Left: Hard components from 0-12\% central 200 GeV \auau\ collisions for identified pions (solid points \cite{hardspec}) and from 200 GeV \pp\ collisions (open points \cite{ppprd}). The dashed and dash-dotted curves show the GLS references for \pp\ and \auau\ collisions. The solid curve is a pQCD description of \auau\ HC data assuming a slight modification of \ee\ fragmentation functions~\cite{fragevo}.
Right: Spectra for identified protons (green curves) from several centralities of \auau\ collisions. The red dotted curve is the TCM soft component. The dashed curve is the TCM proton hard component inferred from \auau\ spectrum systematics. The dash-dotted curves show GLS references for peripheral and central \auau\ collisions. The data hard components for more-central collisions show an increase relative to GLS {\em above} the HC peak mode at $y_t = 2.7$ (1 GeV/c) in contrast to the identified-pion result.
 }  
 \end{figure}

Figure~\ref{aaspec} (left panel) shows \auau\ spectrum HC data for identified pions from 0-12\% central 200 GeV \auau\ collisions (points). Hadron spectrum densities have the form $\rho_{0h} = d^2n_{h}/2\pi y_t dy_t d\eta$. Also shown are the \pp\ TCM HC model $H_{pp}(y_t)$ derived from \pp\ collisions (dashed curve) and the GLS prediction for central \auau\ (dash-dotted curve, 5 times dashed curve).  The centrality-dependent data HC (solid dots)  is obtained by
\bea \label{aatcmeq}
\nu H_{AA}(y_t,\nu) &=& (2/N_{part}) \rho_{0\pi}(y_t,\nu) - S_{NN}(y_t),
\eea
where $S_{NN}(y_t)$ is by definition the limiting case of $(2/N_{part}) \rho_{0h}(y_t,\nu)$ as $\nu \rightarrow 0$ (no dijet production)~\cite{hardspec}. As noted, $S_{NN}(y_t)$ is consistent with the shape of the SC model $\hat S_0(y_t)$ inferred from \pp\ collisions. For the GLS reference $H_{AA}$ retains the fixed form $H_{NN} \approx H_{pp}$ independent of \aa\ centrality and the HC scales with $\nu$.

Above \yt\ = 4 ($p_t \approx 4$ GeV/c) the central-\auau\ pion spectrum is suppressed by factor 5, consistent with conventional $R_{AA}$ measurements, and thus  happens to coincide with the \pp\ hard component. At lower momenta (below \yt\ = 3.3 or $p_t = 2$ GeV/c)  the hard-component data rise far above the GLS reference, which fact is obscured by $R_{AA}$ due to severe bias by inclusion of the spectrum SC in that measure. The solid curve is a pQCD calculation based on a measured MB parton spectrum folded with parametrized FFs~\cite{fragevo}. The same calculation for \pp\ collisions (dashed curve) describes the \pp\ hard-component data (open circles) quantitatively at the 10\% level. For more-central \auau\ collisions the FF parametrization is modified (one parameter is changed by 10\%). The parameter change is equivalent to changing a gluon splitting function in the DGLAP equations~\cite{borg}. The theory description (solid curve) is then accurate over the entire $y_t$ acceptance. On that basis  there can be little doubt that the TCM pion spectrum HC is jet related even in central \auau\ collisions.

Figure~\ref{aaspec} (right panel) shows full-spectrum data for identified protons. The proton spectra (thin solid curves) are compared with inferred spectrum SC $S_{NN}(y_t)$ (dotted curve) common to all centralities. The dashed curve is the \pp-equivalent proton HC model $H_{NN}$ derived as the limiting case of the \auau\ HC for $\nu \rightarrow 1$ (isolated \nn\ collisions). It is remarkable that the proton HC amplitude near $y_t = 2.7$ ($p_t \approx 1$ GeV/c) is comparable to the pion HC amplitude at the same point, although the spectrum SCs are very different~\cite{hardspec}. 

The proton HC evolution with \auau\ centrality is formally similar to that for pions in the sense that strong reduction at larger \yt\ is compensated by enhancement at some smaller \yt, but for protons the enhancement occurs only above 1 GeV/c (approximately the proton mass). The dash-dotted curves marked GLS are defined as $S_{NN} + \nu H_{NN}$ with $\nu = 1.25$, 5.8. The spectrum data for protons in central collisions cross the GLS curve at $y_t \approx 4$ (4 GeV/c), whereas the crossing for pions is at $y_t \approx 3$ (1.3 GeV/c). And below 1 GeV/c the proton HC is consistent with {\em no FF modification}, follows the GLS.  

The difference in evolution between proton and pion HCs fully accounts for the so-called baryon-meson (B/M) puzzle conventionally attributed to CQ coalescence. These results suggest that the B/M puzzle is actually an aspect of parton fragmentation possibly related to hadron fragment mass~\cite{hardspec}. The centrality evolution of proton spectra also strongly supports interpretation in terms of minimum-bias jets.

\subsection{\auau\ number angular correlations}

\begin{figure}[h]
  \includegraphics[width=1.65in]{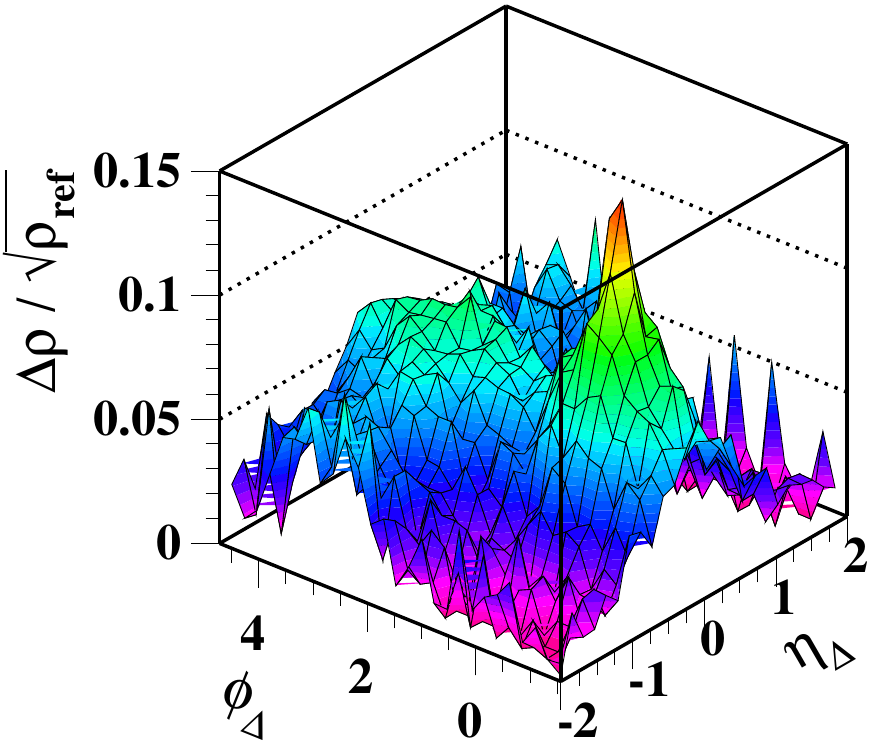}
  \includegraphics[width=1.65in]{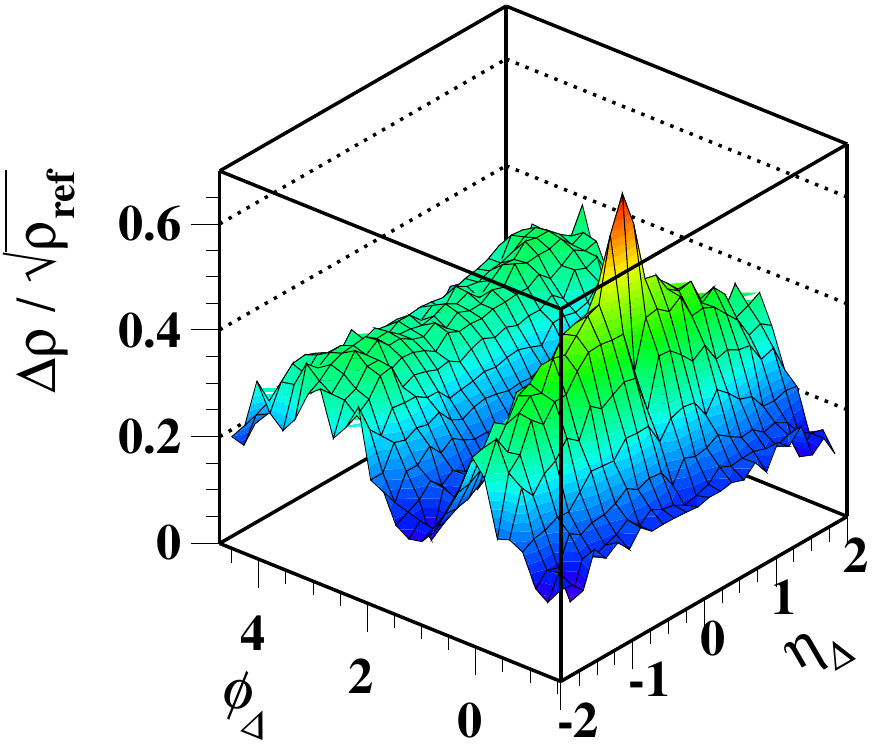}
\caption{\label{aacorr1} (Color online)
Left: Angular correlations from 85-95\% central 200 GeV \auau\ collisions equivalent to NSD \nn\ collisions. The soft component is a 1D peak on $\eta_\Delta$. The hard components are a SS 2D peak and AS 1D peak on azimuth. Conversion electrons and Bose-Einstein correlations contribute a narrow exponential peak at the origin confined to $p_t < 0.5$ GeV/c.
Right: The same for 0-5\% central \auau\ collisions. The SS 2D peak is strongly elongated on $\eta_\Delta$. The soft component falls to zero by mid-centrality. The nonjet quadrupole ($v_2$) amplitude is negligible~\cite{davidhq,davidhq2}.
 }  
 \end{figure}

Figure~\ref{aacorr1} shows 2D number angular correlations from 200 GeV \auau\ collisions for all $p_t > 0.15$ GeV/c and for peripheral (left panel, 85-95\%) and central  (right panel, 0-5\%) collisions. In both cases a SS 2D peak and AS 1D peak on azimuth are evident as for \pp\ collisions. Absent the restrictive \pt\ cuts invoked for Fig.~\ref{ppcorr} (right panel) a narrow 2D peak attributed to conversion-electron pairs and Bose-Einstein correlations (at the origin) and a 1D peak on $\eta_\Delta$ (soft component, left panel only) appear. In the right panel  the SS 2D jet peak in central collisions is strongly elongated on $\eta_\Delta$. A nonjet quadrupole component proportional to $\cos(2 \phi_\Delta)$ is not evident for these centralities but does appear for mid-central data. The z-axis zeros are defined by 2D-model fits to data in which it is assumed that the AS 1D peak is positive definite and the nonjet quadrupole has zero mean~\cite{anomalous}.

\begin{figure}[h]
  \includegraphics[width=1.65in]{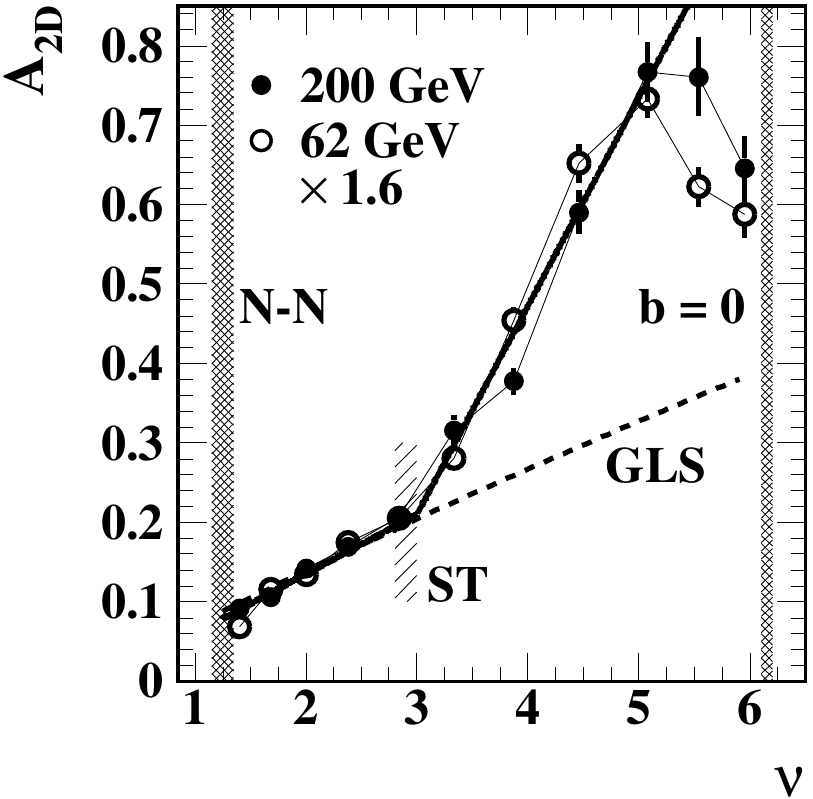}
  \includegraphics[width=1.65in]{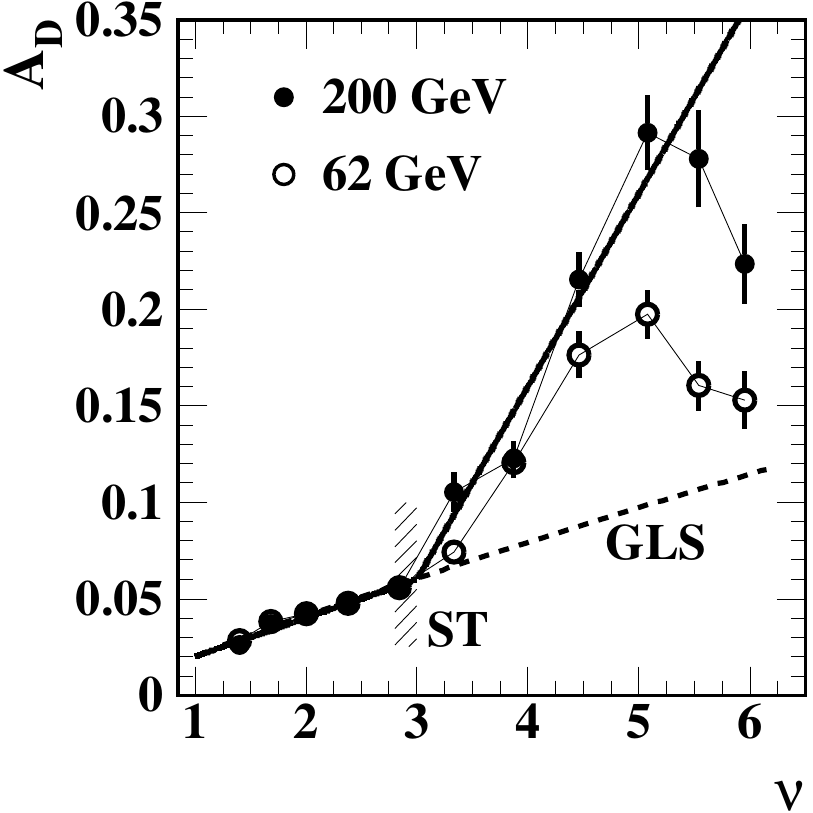}
\caption{\label{aacorr2}
Left: The SS 2D peak amplitude vs centrality measured by pathlength $\nu$ for two collision energies~\cite{anomalous}. The 62.4 GeV data coincide with 200 GeV data when multiplied by factor 1.6, as discussed in the text. The data follow the GLS reference (dashed curve) up to $\nu = 3$ (sharp transition or ST) after which the slope changes by factor 3 or more.
Right: Equivalent results for the AS 1D peak amplitude. The 62 and 200 GeV data correspond for more-peripheral collisions without additional factor. That relation is expected for dijets.
 }  
 \end{figure}

Figure~\ref{aacorr2} shows the centrality dependence of the SS 2D peak (left panel) and AS 1D peak (right panel) amplitudes for 200 GeV (solid points) and 62.4 GeV (open points). The GLS reference for those per-hadron fit-model parameters is the dashed curves defined by $A_{X,pp}\, \nu / [1 + 0.02 (\nu - 1)]$, where the numerator is consistent with the HC from  \pp\ correlations and the denominator is the GLS hadron-yield centrality trend. The data follow the GLS trend up to the ST at $\nu \approx 3$ ($\sigma / \sigma_0 \approx 0.5$). Within a single centrality bin the slopes increase dramatically, and the amplitudes continue to increase rapidly until $\nu \approx 5$ beyond which the data appear to fall off.

In the left panel the 62.4 GeV $A_{2D}$ data multiplied by factor 1.6 coincide with the 200 GeV data. In Ref.~\cite{ptedep} it was observed that the energy dependence of $p_t$ correlation amplitudes attributed to MB jets appeared to vary as $ \log(\sqrt{s_{NN}})$ and to extrapolate to zero amplitude near $\sqrt{s_{NN}} = 10$ GeV suggesting an energy dependence $\approx \log(\sqrt{s_{NN}}/Q_0)$ with energy scale $Q_0 \approx 10$ GeV. The jet-related amplitudes in the left panel are consistent with that conjecture since $\log(200/9) / \log(62.4/9) = 1.6$. The energy trend suggests that dijet production drops to zero near 10 GeV due to kinematic constraints on parton (mainly gluon) fragmentation to charged hadrons~\cite{fragevo}.

The AS peak amplitudes for two energies in the right panel coincide over most centralities with no scale factor. That is expected if the energy dependence is due to the increased kinematic range of colliding small-$x$ partons with increasing collision energy. On 2D dijet rapidity space $(y_{z1},y_{z2})$ a dijet is represented by a single point. 
We speculate that while the kinematic boundaries of that space may expand with increasing collision energy the dijet density remains approximately the same, and the AS 1D peak represents the unchanging 2D dijet density. The SS 2D peak amplitude represents a 1D projection of that space onto its difference axis and therefore does scale with $\log(\sqrt{s})$ changes in  the kinematic boundary on $y_z$.

\subsection{\auau\ $\bf p_t$ angular correlations} \label{ptangle}

The plots in Figs.~\ref{ppcorr} (right) and \ref{aacorr1} represent hadron number  angular correlations. Angular correlations of hadron \pt\  can be obtained either by inversion of the scale dependence of $\langle p_t \rangle$ fluctuations~\cite{ptscale,inverse} or by direct pair counting. The same jet-related correlation structures are observed, with minor quantitative differences in the SS 2D peak structure. Such results indicate that $\langle p_t \rangle$ fluctuations, once expected to reveal critical fluctuations of temperature near a QCD phase boundary, are actually dominated by a MB jet (minijet) contribution. 

\begin{figure}[h]
  \includegraphics[width=1.65in]{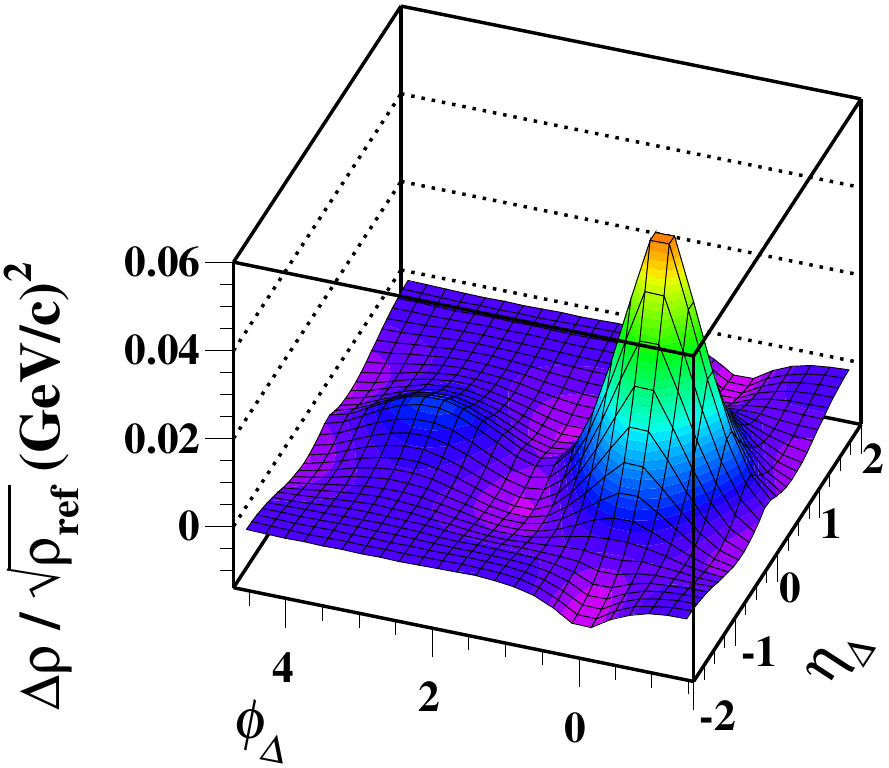}
  \put(-95,90) {\bf (a)}
  \includegraphics[width=1.65in]{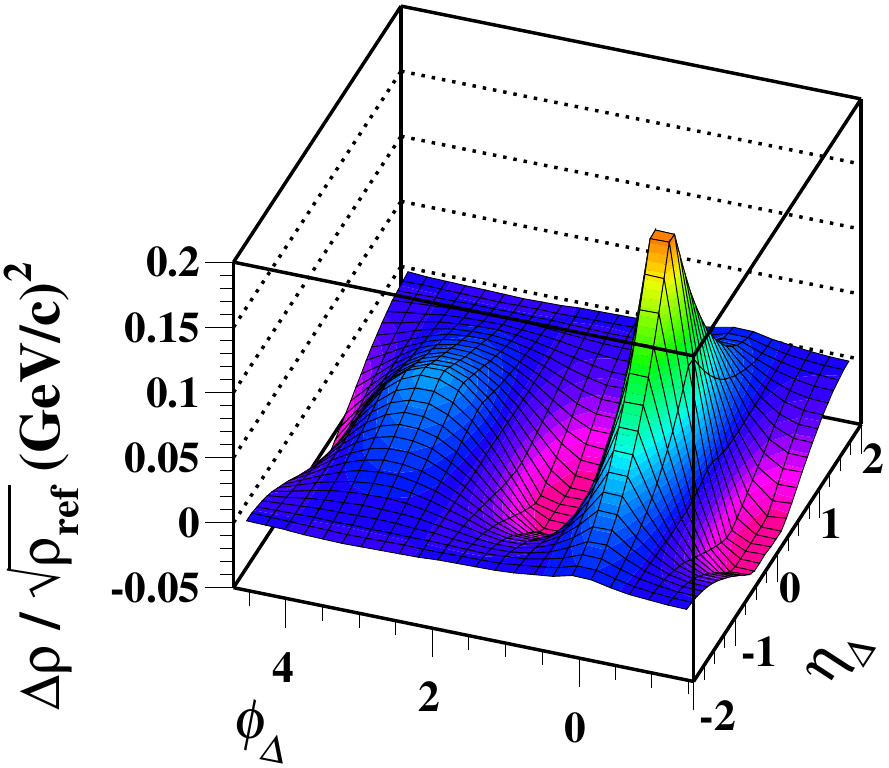}
  \put(-95,90) {\bf (b)}  \\
  \includegraphics[width=1.65in]{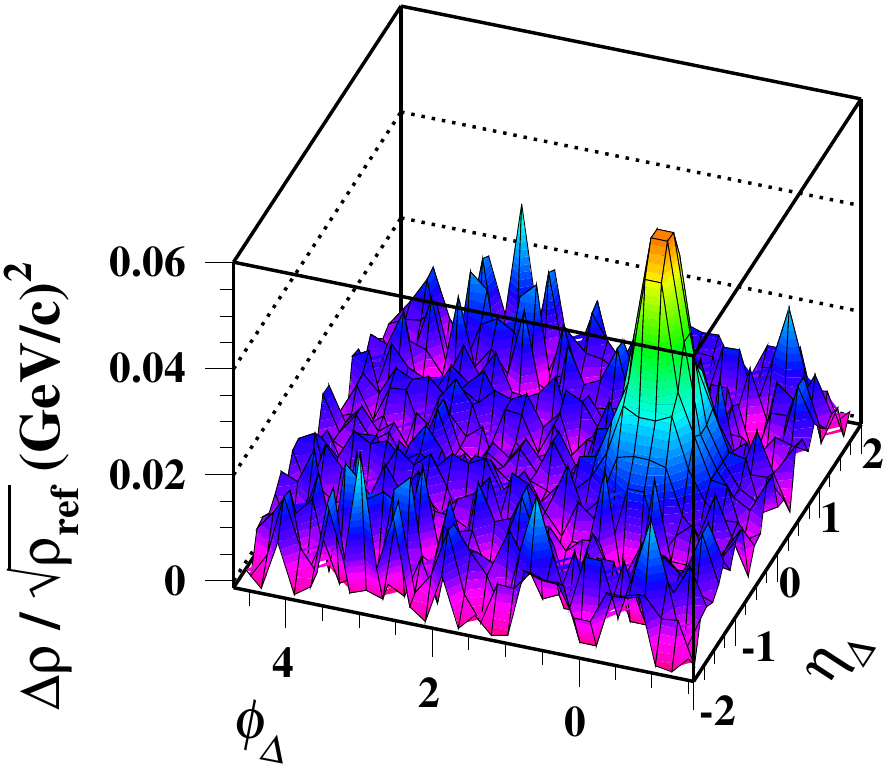}
  \put(-95,90) {\bf (c)}  
  \includegraphics[width=1.65in]{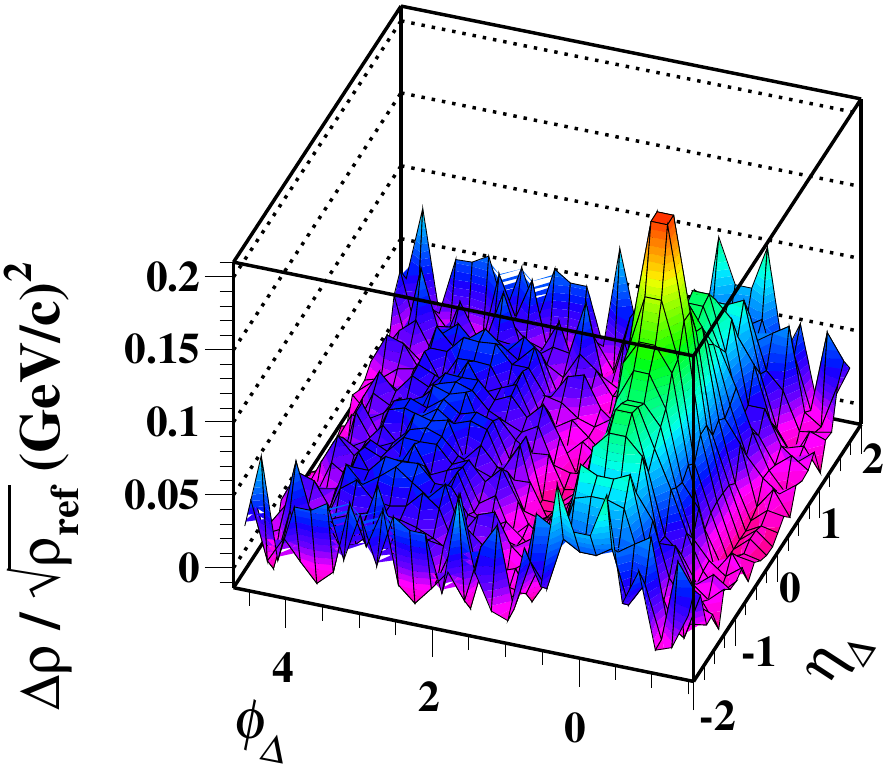}
  \put(-95,90) {\bf (d)}\caption{\label{ptscale} (Color online)
Upper: \pt\ angular correlations for  (a) 85-95\% and (b) 10-20\% central 200 GeV \auau\ collisions inferred by inverting \mpt\ fluctuation scale dependence~\cite{ptscale}. AS dipole and nonjet quadrupole components of 2D model fits to the data have been subtracted.
Lower:  Results for the same collision systems but the \pt\ correlations are obtained by direct pair counting rather than fluctuation inversion. Improved angular resolution is evident as well as unfiltered statistical fluctuations.
 } 
 \end{figure}

Figure~\ref{ptscale} (upper panels) shows \pt\ angular correlations for  (a) 85-95\% and (b) 10-20\% central 200 GeV \auau\ collisions obtained by scale inversion of \mpt\ fluctuations~\cite{inverse}. Such 2D angular correlations are described by a standard fit model including several elements~\cite{ptscale,anomalous}. In this case fitted AS dipole and nonjet quadrupole components have been subtracted to isolate the SS 2D peak structure. A similar analysis of HIJING Monte Carlo data confirms a jet interpretation for that structure~\cite{hijscale}.

Figure~\ref{ptscale} (lower panels) shows  \pt\ angular correlations for the same collision systems obtained by direct pair counting, confirming the results in the upper panels obtained by fluctuation scale inversion. The inversion process smooths the data (regularization): statistical fluctuations are reduced but angular resolution is also reduced. The SS 2D peak for \pt\ correlations is narrower than that observed for number correlations [compare panel (c) with Fig.~\ref{ppcorr} -- right]. That difference is expected for jet correlations, since fewer fragments with larger momenta are found closer to the jet thrust axis and more fragments  with smaller momenta appear at larger angles.

Two features of the SS 2D peak in \pt\ correlations are especially notable. The negative-going region on either side of the peak near the origin suggests the possibility of a recoil component. These are covariance densities, and negative covariance corresponds to anticorrelation of momenta. The other feature is the ``ridge'' especially obvious in panel (d). The strong \pt\ correlations at larger $\eta_\Delta$ argue against $\eta$ elongation of the SS 2D peak in more-central \auau\ collisions residing within a soft background. Related studies of number angular correlations with \pt\ cuts indicate that the SS peak $\eta$ elongation persists up to 4 GeV/c hadron momentum~\cite{davidhq2}, a trend not likely to arise from a soft process.

Thus, in several ways the hard components of 2D number and \pt\ angular correlations from \auau\ collisions follow trends expected for dijets. Dijet production closely follows the predicted GLS trend over half the \auau\ total cross section then undergoes a substantial change within a small centrality interval, described as the sharp transition. Above the ST the correlation structure continues to follow expectations for parton fragmentation to jets, but with fragmentation substantially modified.

\subsection{Correspondence of spectra and correlations} \label{speccorr}

\begin{figure}[h]
  \includegraphics[width=1.65in]{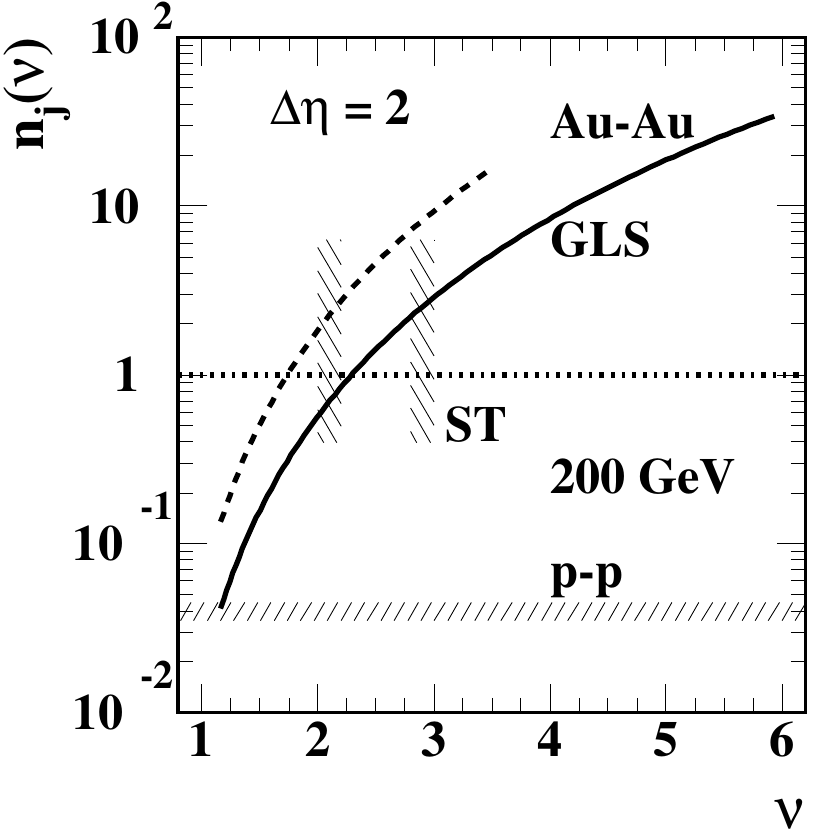}
  \includegraphics[width=1.65in]{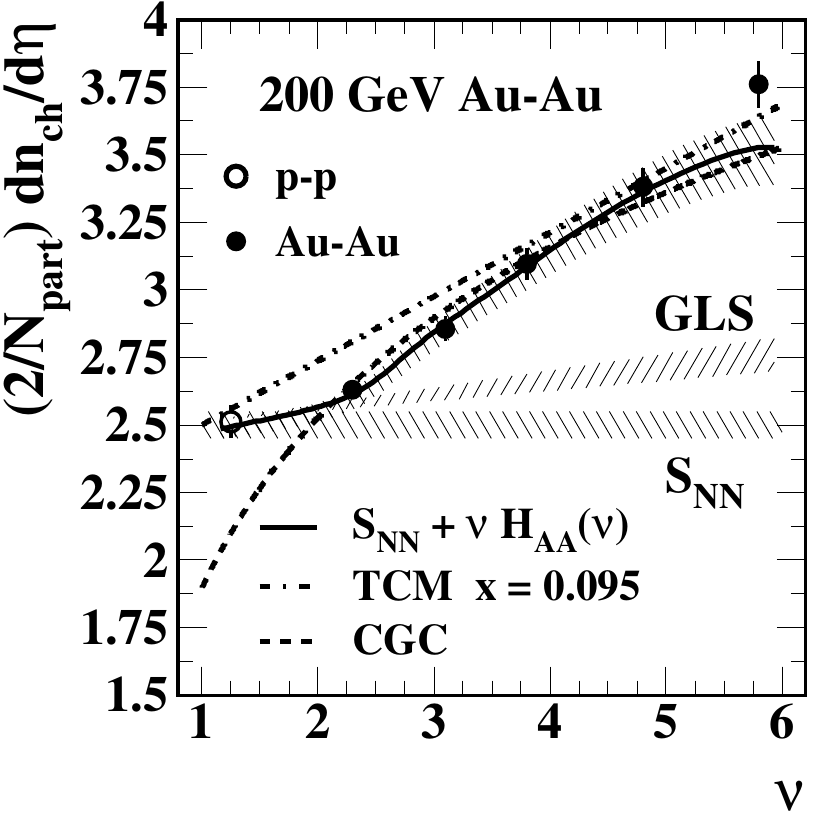}
\caption{\label{corresp}
Left: Dijet number within angular acceptance $\Delta \eta = 2$ at mid-rapidity for 200 GeV \auau\ collisions scaling with $N_{bin}$ from Eq.~(\ref{glauber}) (GLS reference, solid curve). The horizontal hatched band indicates the measured dijet rate for NSD \pp\ collisions in the same acceptance. The upper-right hatched band indicates the sharp transition (ST) near $\nu = 3$ for jet correlations in \auau\ collisions. The dashed curve and left-most hatched band are discussed in App.~\ref{sharpenergy}. 
Right: Per-participant hadron production measured by $(2/N_{part}) dn_{ch}/d\eta$ vs $\nu$ for 200 GeV \auau\ collisions (solid points) inferred from analysis of identified-hadron spectra~\cite{hardspec}. The dash-dotted line is the conventional TCM with fixed $x = 0.095$~\cite{kn}. The solid curve is a prediction obtained by analysis of jet-related angular correlations based on the $n_j$ solid curve in the left panel~\cite{jetspec}. The dashed curve is a prediction derived from color-glass condensate theory~\cite{kn,glasma1}.
 } 
 \end{figure}

Further support for a jet interpretation of the TCM HC is provided by quantitative correspondence between spectrum and correlation data based on a dijet hypothesis. In Fig.~\ref{corresp} (left panel) the solid curve describes the number of dijets (at least one jet) appearing within angular acceptance $\Delta \eta = 2$ (the STAR TPC acceptance) as a function of mean participant-nucleon pathlength $\nu$. The \pp\ value $n_j(NSD) \approx 0.04$ is obtained from Eq.~(\ref{f}). The solid curve is  the \auau\ GLS reference $n_j(\nu) = n_j(NSD)  N_{bin}$. The hatched band labeled ST indicates the position of the 200 GeV \auau\ sharp transition, which happens to correspond to dijet $\eta$ density $f \approx 1$. The dashed curve and left-hand hatched band are discussed in App.~\ref{sharpenergy}.

Figure~\ref{corresp} (right panel)  shows  200 GeV HC yields predicted from jet-related angular correlations summed with a measured soft-component contribution $S_{NN}$ (solid curve~\cite{jetspec}) compared to yields obtained directly from spectra (points~\cite{hardspec}).  The volume of the SS 2D peak in angular correlations is processed using the dijet number per event in the left panel to derive the number of correlated pairs per dijet. The number of correlated pairs is converted to  the number of fragments per dijet and multiplied by $n_j$ to obtain the predicted fragment number per event $\nu H_{AA}$. When that HC is added to SC $S_{NN}$ we obtain the TCM prediction for charged-hadron yields. The agreement with integrated yields from hadron spectra (points) is within systematic uncertainties (upper hatched band). The dash-dotted line shows the conventional 200 GeV \auau\ TCM with fixed $x = 0.095$~\cite{kn}. The lower hatched band is the GLS reference TCM with $x = 0.015$ derived from \pp\ spectrum data. The dashed curve shows a prediction based on CGC theory~\cite{glasma1,kn}. This comparison provides additional strong support for a dijet interpretation of the HC of spectra and correlations for all \auau\ centralities.

\section{Challenging  the CQM} \label{response}

We now confront the conjectured CQM with experimental evidence summarized in the previous two sections and arguments based on the historical development of QCD and logical structure.

\subsection{Arguments from historical concepts}

Some supporting arguments for the CQM appear to be based on historical concepts of hadron production in HE nuclear collisions. It is notable that in Ref.~\cite{phenix} 31 references appear after 1990 whereas 48 appear before that date [note the relation to LEP commissioned in 1989 and the Hadron-Electron Ring Accelerator (HERA) commissioned in 1992]. The HERA \ep\ collider experiments provided a large volume of accurate data on hadron (proton) internal structure, including parton distribution functions (PDFs).  The LEP \ee\ collider experiments provided a large volume of accurate data on parton FFs over a broad range of dijet energies and parton types. The two data volumes, combined with pQCD theory, accurately and comprehensively describe hadron production in HE nuclear collisions, including dijets~\cite{eeprd,fragevo}.

Historical issues can be considered in a context where the full $4\pi$ solid angle is divided into a limiting-fragmentation (LF) region near the beam rapidity and a mid-rapidity (MR) region near the collision center of momentum. With the discovery of partons (mainly valence quarks) in lower-energy \ep\ collisions at SLAC and the initial development of QCD the search commenced for hard parton scattering as a transport process from LF to MR in HE collisions. Hard parton scattering to jets should occur rarely, and the evidence should appear as transport from LF to MR. Jet-related hadron production in the MR should be sparse and highly structured.

As recounted in Ref.~\cite{phenix}, contrary to expectations copious \et\ and hadron production was actually observed in the MR. However, the production appeared to be weakly structured on angle and approximately  thermal in  terms of the fixed relation of \et\ to hadron  multiplicity. That unexpected result lead to the conclusion that production in the MR must be dominated by soft processes such as QCD string fragmentation~\cite{lund}, with negligible jet contributions. The strings were thought to connect colored valence quarks/diquarks  or dressed constituent  quarks (valons~\cite{valons}). That historical context apparently prompted Ref.~\cite{phenix} to the conclusion that inference of copious (semi)hard scattering to the MR (minijets) ``seems to contradict the extensive measurements...[of \et\ and $n_{ch}$]...which show that  these distributions represent measurements of the `soft' multiparticle physics that dominates...'' \pp\ collisions. As a consequence, the study in Ref.~\cite{phenix} is restricted to ``...the nucleon and constituent quark participant models of soft-multiparticle production widely used since the 1970's....''

\subsection{Arguments from p-p phenomenology} \label{pparg}

The importance of projectile small-$x$ partons (mainly gluons) to hadron production in  the MR has already been mentioned in Sec.~\ref{lowxglue}. Here we review some detailed aspects of \pp\ data phenomenology pertaining to that issue. The CQM must contend with  the accurately-measured $n_{ch}$ dependence of \pp\ collisions and copious dijet production for larger event multiplicities. Aside from substantial MB dijet production at mid-rapidity another major issue is the dominant role of fluctuations.

The integrated MR charge $\rightarrow n_{ch}$ in 200 GeV \pp\ collisions fluctuates event-wise over a large interval (at least a factor ten), which is problematic  for a CQM with 2-3 participant quarks $N_{qp}$ per projectile and hadrons dominated by string fragments proportional to that number. The resolved-dijet production scales approximately as $n_{ch}^2$, leading to a factor 100 variation in dijet number. For 200 GeV \pp\ collisions with $dn_{ch} / d\eta \approx 25$ ($10 \times$NSD) the mean number of dijets within $\Delta \eta = 2$ is approximately two {\em per event}, and fragments from resolved jets comprise 10\% of all hadrons. A string/CQ model cannot generate dijets at that level (or at all) and cannot describe the large range of multiplicity fluctuations, whereas measured projectile small-$x$ structure and corresponding dijet production based on perturbative QCD can do so given event-wise fluctuations in the parton splitting cascade within a proton~\cite{pptheory}.

Figure~10 of Ref.~\cite{phenix} provides another indicator. The gamma-distribution model function representing the CQM (solid curve) deviates strongly from the tail of the \pp\ data distribution that represents large fluctuations in \et\ production. When applied in Eq.~(7) to construct the CQ MB model function as a convolution the result actually fails to represent the MB data. Close examination of Fig.~11 at the terminus reveals that the CQM (red curve) has a much larger slope (smaller fluctuation width) than the data (blue points). The apparent agreement of the MB distributions arises directly from the relation shown in Fig.~\ref{consquark} (right panel) of the present study. The substantial deviations for more-peripheral collisions that may falsify the CQM are not visible in the conventional (and insensitive) semilog plotting format.

Use of a Glauber Monte Carlo to estimate the \pp\ $N_{qp}$ as in Fig.~\ref{glauber} seems problematic. In Sec.~\ref{ppmb} we note that the trend $N_{bin} \propto N_{part}^2$ (with $N_{part} \sim n_s$) that describes dijet production in \pp\ collisions in terms of small-$x$ parton participants is inconsistent with the eikonal approximation which corresponds to  $N_{bin} \propto N_{part}^{4/3}$. But the Glauber model is based on the eikonal approximation and cannot therefore be used  to model \pp\ collisions in terms of conjectured CQ participants. The value $N_{qp} = 2.8$ inferred from a \pp\ Glauber simulation is inconsistent with the value $N_{qp} \approx 4$ obtained in Fig.~\ref{glauber} by extrapolating the \auau\ trend (where the eikonal approximation and Glauber model are valid) to \pp\ collisions for $\nu \approx 1$. That is a significant issue because the value $N_{qp} = 2.8$ is supposed to indicate self-consistency of the CQM as in Sec.~VIII-B of Ref.~\cite{phenix}.

\subsection{Arguments from \aa\ phenomenology} \label{aaarg}

The CQM appears to be relevant because the simulated $N_{qp}$ centrality trend approximates the \aa\ centrality trends for $n_{ch}$ and \et, at least for more-central collisions. However, major problems arise for $N_{qp}$ from more-differential analysis. 
Below the sharp transition near $\nu = 3$  \auau\ spectra and correlations follow a TCM reference (GLS) quantitatively equivalent to that describing  \pp\ collisions, with $n_s$ replaced by $N_{part}$ as the SC control parameter and $n_h / n_s$ replaced by $\nu$ as a measure of \nn\ binary collisions~\cite{anomalous}. The number of resolved dijets $n_j$ having correlation structure consistent with in-vacuum pQCD jets increases from 0.04 per event to about  2 per event within the STAR TPC acceptance, scaling with the number of \nn\ binary collisions as expected for dijets. The same jet correlation structure appears in both number and \pt\ angular correlations. Substantial deviations from the $N_{qp}$ trend for more-peripheral collisions are already evident in Fig.~\ref{consquark} (right panel). If the CQM is excluded by the dijet presence in \pp\ collisions it is excluded in more-peripheral \auau: the spectrum HC remains quantitatively related to pQCD dijets. 

In more-central collisions above the sharp transition the dijet contribution is altered quantitatively, but the spectrum HC is still described by pQCD modulo a simple modification of FFs that conserves the parton energy and remains consistent with the general form of the DGLAP equations. The same basic correlation structure persists even in central collisions and still scales with $N_{bin}$, with {\em no actual reduction from the expected dijet number}. In central 200 GeV \auau\ collisions we find that 1/3 of the final state is included within resolved dijets~\cite{jetspec,anomalous}, dramatically inconsistent with the CQM

Reference~\cite{phenix} states that the form of $(2/N_{part}) dn_{ch}/d\eta$ vs $N_{part}$ is statistically equivalent at 200 GeV and 2.76 TeV (ALICE data) ``although the jet cross section increases by a very large factor.'' The conclusion is drawn that hadron production must then be dominated by soft processes with no significant dijet contribution. But careful examination of the comparison reveals several issues, as established in App.~\ref{twoenergies}. The shape of the centrality trend for PHENIX data is substantially distorted by fluctuations for more-central collisions due to the small angular acceptance. The sharp transition at 200 GeV near $\nu = 3$ appears to be significantly lower at 2.76 GeV. When  the comparison is made by scaling the ALICE data down by the same factor 2.1 the more-peripheral 2.76 TeV data fall well below the measured 200 GeV NSD \pp\ numbers. Thus, a combination of several mitigating factors may conspire in the original comparison to give a false impression. We find that a TCM scaled up by factors 1.8 and 1.8$^2$ derived from RHIC energy systematics below 200 GeV predicts the ALICE data trend within rather small data uncertainties. The TCM is actually strongly supported by that comparison.

Reference~\cite{phenix} argues that dijets are strongly suppressed in more-central \auau\ collisions (jet quenching). So a TCM increase in hadron production due to dijets would be inconsistent with jet suppression inferred from $R_{AA}$ data. However, spectrum data presented in Sec.~\ref{aaspectra} reveal that the actual jet-related hadron production integrated over the full \pt\ acceptance {\em increases} with centrality by up to a factor six. The noted suppression at higher \pt\ is compensated by a much larger enhancement at lower \pt\ (so as to conserve the parton energy), still within the collimated jet structure~\cite{hardspec,fragevo}. That conclusion is supported by jet-related angular correlation trends~\cite{anomalous}.

\subsection{Arguments from p-A phenomenology}

Several lessons from p-A collision data that appear to support the TCM are drawn in Ref.~\cite{phenix}. A projectile nucleon (participant) interacting with a target nucleus may be excited (wounded) only once and thereafter does not change its internal state in subsequent \nn\ collisions. Because of time dilation  the excited nucleon cannot fragment within the target nucleus. The projectile (p) hemisphere is independent of A, and the target (A) hemisphere depends on $\nu$,  the mean path length of the projectile in the target. The formation-time argument appears to eliminate the possibility for  rescattering of secondary partons or hadrons: ``This feature [formation time] immediately eliminates the possibility of  a cascade in the nucleus from rescattering of the secondary products''~\cite{phenix}.

Certain implications follow for \aa\ collisions. Absence of rescattering would explain why 3 GeV minijets survive intact without significant loss in central \auau\ collisions~\cite{anomalous}. It would also explain the absence of collective (radial and elliptic) flows indicated by recent observations~\cite{hardspec,davidhq,davidhq2}. Soft hadron production by linearly-independent projectile-nucleon dissociation would then scale as $N_{part}$ with properties independent of \aa\ centrality. Dijet production would scale as the number of \nn\ binary collisions $N_{bin}$. And the number of dijets per participant nucleon would scale with  $\nu$ adopted from the p-A phenomenology. But that is exactly the \aa\ TCM.

\subsection{Arguments from logical structure}

Reference~\cite{phenix} seeks to verify what may be called hypothesis A: Almost all hadron and transverse momentum/energy production in the MR results from soft hadron production following $N_{qp}$ scaling. The paper establishes that A implies certain data trends B: (i) $(2/N_{qp}) dE_t/ d\eta$ should be approximately constant with \auau\ centrality and (ii) the MB distribution on $N_{qp}$ may be used to generate a model distribution on \et\ that should  describe MB data. Evidence apparently supporting B is provided and it is concluded that hypothesis A is valid. We note that (i) and (ii) are mathematically equivalent, and results B rely on an approximate relation between $N_{qp}$ and $dn_{ch}/d\eta$ that may be accidental.

The logical structure of such an argument is known to be questionable. Other hypotheses A$'$ may also imply B, and other results B$'$ may falsify A. Observation of \=B would indeed imply \=A, falsification of the hypothesis. But observation of B does not imply or require A, nor does it falsify alternatives A$'$ (e.g., the TCM). Examples of phenomena B$'$ that falsify A include the $n_{ch}$ systematics of dijet production in \pp\ collisions, agreement with the GLS trend below the ST in \auau\ collisions, strong jet-related correlations for all collision systems, and the central role of $N_{bin}$ HC scaling in all collision systems.

In contrast, the TCM is implied by modern QCD theory combined with the measured properties of hadrons (PDFs from HERA) and dijets (FFs from LEP). The TCM is required by the phenomenology of \pp\ and \aa\ collisions, including differential spectrum and correlation measurements. The TCM is not an empirical description of data (as conventionally represented). It is a quantitative prediction based on QCD theory and data  from elementary collisions in the form of the GLS reference. In contrast, the CQM makes no such quantitative predictions and fails to describe data trends for more-peripheral \aa\ collisions. The CQM assumes no hard scattering, so copious evidence for hard scattering from spectra and correlations appears to falsify the CQM.

\subsection{Summary}

With the premise that hadron production in the MR must be controlled by large-$x$ degrees of freedom in the LF represented by conjectured constituent quarks, that hadron production from jets must be sharply reduced with increasing \aa\ centrality as implied by $R_{AA}$ data, and that ``possible models motivated by the fact that half of the momentum of a nucleon is carried by gluons when probed at high $Q^2$ in hard-scattering are not considered,'' the CQM seems to be the only alternative. But the description of high-energy nuclear collisions during the past twenty years has become increasingly reliant on just the small-$x$ partons excluded from the CQM. And the improvement in analysis methods applicable to dijet structure since first RHIC operation is substantial.

We do observe that a soft component compatible with projectile nucleon dissociation plays a major role in hadron production complementary to dijet production, but the SC is easily described by a few universal parameters, with form independent of \aa\ centrality. Without a dijet contribution the CQM must rely on string fragmentation as the sole transport mechanism from LF to MR. The CQ concept and string fragmentation are nonperturbative, and the CQM can thus make no connection to pQCD  theory. Direct evidence from several differential methods for copious dijet production in the MR, described quantitatively by pQCD for all \aa\ centralities, then appears to exclude the CQM. In contrast,  the TCM describes a large variety of data at the few-percent level over a broad range of collision energies from SPS to LHC.

\section{Discussion}\label{disc}

In this section we consider further aspects of hadron production mechanisms and methods intended to reveal them, including measure sensitivity, model energy dependence, the utility of MB distribution modeling and the role of fluctuations in MB distributions and production centrality trends.

\subsection{Sensitivity to hadron and $\bf E_t$ production}

The question posed in the title of the present paper relates to the sensitivity of certain aspects of data to collision mechanisms. What mechanisms (or models thereof) can be tested by what aspects of the data. The ability to test models relies on the amount of information in the data, which in turn depends strongly on how the data are presented (statistical analysis and plotting format).

The greatest information is carried by multiparticle correlations within narrow multiplicity or centrality bins. Successive integrations (and therefore information loss) lead to two-particle correlations (a limiting case), fluctuations (integrals of correlations), spectra, full MB distributions and yields within centrality bins. The information carried by MB distributions is small compared to that in differential 2D correlations and 1D spectra.

Aside from the inherent information content of the data the choice of plotting format can determine the fraction of such information that is visually accessible. The peripheral centrality region is conventionally disregarded, by the choice of centrality interval for actual measurements (the top 40, 65 or 80\% are common choices) and plotting format ($N_{part}$ de-emphasizes the peripheral region). In contrast, a per-participant measure plotted vs $\nu$ provides an ideal format for testing the TCM hypothesis. 

Use of \pt\ instead of \yt\ de-emphasizes the low-\pt\ region where major jet-related variations with \aa\ centrality occur. Spectrum ratio $R_{AA}$ nominally measuring jet structure is insensitive to jets below 4 GeV/c because of the spectrum SC which dominates the ratio below that point, whereas TCM analysis of spectra reveals large HC contributions extending down to 0.5 GeV/c or lower. 
MB distributions plotted in a semilog format on \et, $N_{part}$ or $n_{ch}$ convey {\em no significant information} visually. The same data plotted in a power-law format over the full centrality range convey all available information, including variation of TCM parameter $x$ with centrality and fluctuations for central ($b = 0$) \aa\ collisions and possibly \pp\ collisions.

In Ref.~\cite{phenix} the key data plots are Figs.~3, 5, 6, 7 and 16, with 3 and 16 being semilog plots vs \et. Such data are said to ``provide excellent characterization of the nuclear geometry...and are sensitive to the underlying reaction dynamics....'' But that claim is not born out by the \et\ data. Figs. 5 and 6 on $N_{part}$ are replotted on $\nu$ as Figs.~\ref{mbet} and \ref{met} of the present paper. (Fig.~7 is effectively an integral of Fig.~6 and as such carries less information.) Whereas the ST is a prominent feature of the centrality dependence of correlations, spectra and integrated hadron yields from \auau\ collisions consistent with the TCM and QCD, there is no evidence for the ST in the \et\ data as plotted, no sensitivity to that important new phenomenon. Part of the problem is the large systematic uncertainties, especially for more-peripheral collisions, presumably arising from uncertainty in collision centrality and therefore correct values for $N_{part}$ or $N_{qp}$. Thus, although the TCM and CQM strongly disagree in more-peripheral collisions the data presented in Ref.~\cite{phenix} cannot test the model differences.

The composite nature of \et\ presents another issue. The \et\ integrated within some acceptance can be factored into the integrated hadron multiplicity $n_{ch}$ and mean \et\ per hadron denoted by \meet. A  similar argument pertains to $\langle p_t \rangle$ fluctuations~\cite{ptfluct}. Given the information extracted from $n_{ch}$ distributions we seek unique aspects of \et\ production beyond hadron multiplicities. Such information may be carried in per-hadron mean values as shown in Fig.~\ref{met}. Unfortunately, the large systematic uncertainties preclude effective model tests, especially regarding detailed structure predicted by the TCM and accessed by differential spectrum analysis. One may conclude that in the CQM context there is no new information from \et\ measurements beyond  hadron multiplicities.

\subsection{$\bf \sqrt{s_{NN}}$ dependence of hadron and $\bf E_t$ production}

The energy comparison presented in App.~\ref{twoenergies} has interesting implications. The production data at 2.76 TeV are described accurately by a TCM derived from 200 GeV data with two adjustments: The soft component is multiplied by factor 1.8, the hard component by factor $1.8^2 = 3.24$. The factor 1.8 is inferred from a $\log(\sqrt{s}/Q_0)$ trend (with $Q_0 \approx 9$ GeV) inferred from RHIC data below 200 GeV and interpreted to represent production of participant small-$x$ gluons (soft component) in \pp\ or \nn\ collisions which may then collide as binary pairs to produce dijets (hard component). The result in Fig.~\ref{alice} (left panel) suggests that modifications to jet formation in \pbpb\ collisions at 2.76 TeV are {\em essentially identical} to those in \auau\ collisions at 62.4 and 200 GeV. The only difference may be a shift of the ST to more-peripheral collisions. Substantial FF modifications above the ST observed at RHIC energies apparently remain unchanged.

\subsection{Utility of MB distribution modeling}

Reference~\cite{phenix} models the MB distribution on \et\ based on convolution of a model for the \et\ distribution from \pp\ collisions and a MB distribution on $N_{qp}$ derived from a Glauber simulation. The procedure is described as the Extreme Independent Model. The model is compared to data in Fig.~11 and ``excellent agreement'' is reported.  However, one can question the utility of such a procedure.

Equation~(7) of Ref.~\cite{phenix} represents a convolution integral that gives a MB distribution on \et\ given conditional distributions on \et\ for a given number of $N_x$ collisions and a MB distribution on $N_x$, with the form
\bea
\frac{d\sigma}{dE_t} &=& \sum_{N_x} \frac{d\sigma}{dN_x} P(E_t|N_x)
\eea 
where the $P(E_t|N_x)$ are derived by convoluting gamma distributions fitted to \pp\ data. In  the CQM the implicit assumption is invoked that hadron and \et\ production remain the same for all \aa\ centralities. Only $N_{qp}$ varies.

There is an exact equivalence between a MB frequency distribution on quantity X integrated within some angular acceptance and the centrality variation of the produced X. A running integral of the MB distribution gives $\sigma(X)$ vs $X$. Combined with $N_{part}/2$ and $\nu$ vs $\sigma$ from a Glauber MC as in Eq.~(\ref{glauber}) the running integral gives $(2/N_{part}) X$ vs $\nu$. That procedure is used in App.~\ref{tailflucts} to determine the distortion effect of fluctuations at the terminus of the MB distribution on the per-participant production centrality trend, as in Fig.~\ref{flucts} (right panel).

$N_{qp}$ is observed to have approximately the same trend as $n_{ch}$ assuming fixed TCM parameter $x \approx 0.1$ as in Fig.~\ref{consquark} (right panel). Thus, the MB distribution on $N_{qp}$ is equivalent to that on $n_{ch}$ within a rescaling of the $x$ axis.
In App.~\ref{joint} we obtain a correspondence between MB distributions $d\sigma/dE_t \approx (dE_t/dN_{qp})^{-1} d\sigma/dN_{qp}$, where Jacobian $dE_t/dN_{qp} \approx 2/3$ GeV is approximately independent of $N_{qp}$ as indicated in Fig.~\ref{mbet} (right panel). The same argument holds for the relation between $n_{ch}$ and $N_{qp}$, where the Jacobian is $dN_{qp}/dn_{ch} \approx 1.5$. Finally, we have $dE_t / dn_{ch} \approx 1$ GeV to close the circle.

Thus,  the form of  the CQM MB distribution on \et\ is determined to good approximation by that on $n_{ch}$ and the equivalent on $N_{qp}$. Any MB distribution within that ``family'' can be generated from the TCM by the method described in App.~\ref{mbstruct}. The only difference is the Jacobian factor. The terminus half-maximum values are then related as follows (referring to figures in Ref.~\cite{phenix}): $N_{part}/2 = 191$ [Eq.~(\ref{glauber})], $N_{qp} = 2\times 2.8 \times 191 = 1070$ (Fig.~9) $E_t = 0.66 \times 1070 = 710$ (Fig.~11 -- upper panel).
The comparison between CQM distribution and data in Fig.~11 of Ref.~\cite{phenix} is said to indicate ``excellent agreement,'' including the matched terminus positions. But the general MB model shape {\em must} describe the  data as implied by Fig.~\ref{consquark} (right panel), and the endpoints are determined by Jacobians already established elsewhere. 

The only unique information in a MB distribution within this family of kinematic quantities, beyond their production trends on centrality, is the slope at the terminus representing fluctuations in central collisions. Close examination of the terminus shapes in Fig.~11 reveals substantial disagreement. The model slope ($\propto 1/\sigma_{E_t}$) is much larger than the data, indicating that fluctuations are underestimated.  But the CQM assumes the same production mechanisms for \pp\ collisions and central \auau\ collisions, whereas the correlation structure in the two cases is very different~\cite{anomalous}, and fluctuations reflect (are integrals of) that correlation structure~\cite{inverse}. 

There are substantial differences between the CQM and the TCM (and the data it describes accurately), but those differences (reflected for instance by significant variation of the Jacobians noted above) are concealed by the insensitive semilog format invoked for conventional MB distribution plots. Thus, we can conclude that the Extreme Independent Model of Ref.~\cite{phenix} does not reveal additional information beyond what is accessible from production centrality trends, visually suppresses significant model differences within the semilog plotting format and cannot describe fluctuations in central \auau\ collisions. Its utility is therefore questionable.

\subsection{The TCM MB distribution in a CQM context} \label{cqmtcm}

In Sec.~\ref{tcmcq} an argument is presented that the TCM fails to model MB data distributions and should therefore be rejected. However, the TCM implementation is incorrect as demonstrated in this subsection. The general problem relates to the 3D space $(X,N_x,N_y)$ where $X$ is some produced quantity such as \et\ and $N_x$, $N_y$ are parameters modeling \aa\ geometry. The MB density on $(X,N_x,N_y)$ may be normalized to the total cross section $\sigma_0$ or  to unity. For this discussion we choose the latter and discuss probabilities in terms of $P = \sigma / \sigma_0$.

The 3D space can be projected to three 1D marginal spaces. One marginal space $dP/dX$ is directly measurable and the other two are approximated by Glauber simulations. The running integrals in those spaces provide parametric relations, e.g.\ $X(P)$, $N_x(P)$ and $N_y(P)$. Mean values of the three quantities are parametrically related through common parameter $P$ to define a curve or locus of means in  the 3D space. Those relations provide the basis for inferred production centrality trends such as shown in Fig.~\ref{corresp} (right panel). We choose one model parameter $N_x \rightarrow N_{part}$ as the basic degree of freedom for the TCM. We empirically observe simple  power-law trends such as $N_{part}^{1/4}(P) \propto P$ as expressed in Eqs.~(\ref{glauber}). Given that framework we consider details of the TCM as manifested in different contexts.

The MB distribution on $X$ can be related to the MB distribution on one of  the geometry parameters through a Jacobian derived from the parametric relations, as in Eq.~(\ref{jacobeq}). That representation is advantageous because of the simple power-law structures on $N_{part}$ and $N_{bin}$ and the TCM formulation of the parametric relations as in Eq.~(\ref{aatcmeq}). The result is the correct form of the TCM for the MB distribution on $X$ as described in App.~\ref{etdata}, with direct comparisons to data. We find that the TCM accurately describes all the experimentally accessible features of MB data distributions to their uncertainty limits.

The conjectured relations in Eq.~(\ref{phenixmbeq}) represent an alternative implementation of the TCM for MB distributions. Dividing both sides by $\sigma_0$ we obtain probability densities with some interesting features. First, the density in one marginal space is related to a sum of densities in two other marginal spaces. In terms of probabilities that implies a dichotomy: In each event either participants {\em or} binary collisions are related to $E_t$. Second, the Jacobians relating $E_t$ to $N_{part}$ and $N_{bin}$ are both assumed to be equal to fixed value $\langle E_t \rangle^{pp}$. Neither aspect seems plausible. We conclude that Eq.~(\ref{phenixmbeq}) is an improper implementation of the TCM for MB distributions, and TCM comparisons with MB data as in Ref.~\cite{phenix} are misleading.

\subsection{Multiplicity and $\bf E_t$ fluctuations}

Fluctuations in $n_{ch}$, \et\ and other kinematic quantities may convey some information about collision mechanisms. In general, fluctuations represent correlation structure integrated over some angular acceptance~\cite{inverse}. Thus, direct study of correlations is preferred. However, in case of sparse data (rare particles, low event numbers) an integral fluctuation measure may be the only option.

Fluctuations can be measured by specific statistics (e.g., cumulants) or by parametrized model functions fitted to frequency distributions. The negative-binomial distribution (NBD) describes distributions on discrete variables such as $n_{ch}$ whereas the gamma distribution describes distributions on continuous variables such as \et. In either case the limiting form is the Poisson distribution (describing uncorrelated samples).

The NBD has two parameters $\bar n$ and $k$ that are related to a variance (cumulant) by
\bea
  \frac{\sigma^2_n}{\bar n}& =& 1 + \frac{\bar n}{k}.
\eea
In the limit $1/k \rightarrow 0$ the NBD goes to a Poisson distribution. The NBD is a  generic model for fluctuations from any correlated discrete system that may include complex correlation structures (e.g., from dijets) hidden by the integral measure but accessible through other more-differential methods. The expression on  the right represents the leading terms in a power series representing more-complex frequency distributions (e.g., excess events in the distribution tail, higher-order cumulants).

Multiplicity fluctuations are generally biased by fluctuations in the \aa\ collision geometry (width of a centrality bin) that cannot be fully controlled by an external parameter. However, fluctuations for central ($b = 0$) collisions can be estimated from the slope of the terminus of the MB distribution plotted in the power-law format.

In the case of large fluctuation amplitudes the gamma and NBD models may fail to describe data distributions even approximately. For instance, in Fig.~10 of Ref.~\cite{phenix} the p-p MB distribution on \et\ (points) deviates strongly from the gamma model function (solid curves) in the tail region, presumably due to large fluctuations from small-$x$ gluons. For that reason among others the Au-Au CQM in Fig.~11 fails to describe the terminus (the slope is much too large). Further discussion appears in App.~\ref{tailflucts}.

\section{Summary}\label{summ}

Based on comparisons of Monte Carlo simulations for the number of constituent-quark participants $N_{qp}$ with the measured centrality trends of integrated multiplicity $n_{ch}$ and transverse energy \et\ it is argued that the two-component (soft+hard) model (TCM) of mid-rapidity production, including a substantial dijet contribution, is actually a proxy for a constituent-quark (CQ) model (CQM) in which dijets play no significant role. Further support for the CQM is derived from comparison of an Extreme Independent Model of minimum-bias (MB) distributions based on $N_{qp}$ with MB data and comparisons of RHIC and LHC production trends that seem very similar despite  the large energy difference. Hadron and \et\ production near mid-rapidity is seen as arising from fragmentation of QCD color strings joining CQs residing at large momentum fraction $x$ within projectile nucleons.

In the present study we confront arguments supporting the CQM with contrasting differential evidence from yields, spectra and correlations that provide strong support for the TCM and the major role played by dijets in \pp\ and \aa\ collisions near mid-rapidity.  The TCM relates directly to pQCD predictions and provides detailed quantitative descriptions of a broad range of phenomena in correlations, spectra, yields and MB distributions.


Differential spectrum and correlation measurements from 200 GeV \pp\ collisions reveal a dijet contribution quantitatively consistent with pQCD predictions based on measured jet spectra and fragmentation functions, the production depending on the number of small-$x$ partons (mainly gluons) represented by a soft multiplicity component $n_s$. The dijet rate (hard component) is proportional to $n_s^2$ representing parton-parton binary collisions and indicating that the eikonal approximation is not valid for \pp\ collisions. Fluctuations in the soft multiplicity over a ten-fold range correspond to variations in the dijet rate over a hundred-fold range, including multiple dijets per collision at the upper end. The 
\pp\ phenomenology provides the reference for a TCM in \aa\ collisions representing Glauber linear superposition of \nn\ collisions (GLS).

In \aa\ collisions the role of soft multiplicity $n_s$ is assumed by participant-nucleon number $N_{part}$, and \nn\ binary collisions are represented by $N_{bin}$ with ratio $n_h / n_s$ replaced by $\nu = 2N_{bin} / N_{part}$. The \aa\ TCM based on those parameters is observed to follow the GLS reference over half the \auau\ total cross section for 62.4 and 200 GeV collisions, with dijet structure as in \pp\ collisions.
In more-central \auau\ collisions, above a sharp transition in the hard-component centrality trend, the dijet structure is observed to change quantitatively but remains consistent with a slightly-modified pQCD description. Although some dijet properties change the number of dijets remains proportional to $N_{bin}$. No dijets are lost.

The measured energy dependence of the TCM for \auau\ collisions below 200 GeV can be used to predict data trends at LHC energies. The hadron production vs centrality for \pbpb\ at 2.76 TeV is described within small data uncertainties by the extrapolated TCM.

In contrast the CQM is excluded by data in several ways. The \pp\ multiplicity dependence of spectrum and correlation structure, which clearly requires a dijet contribution, cannot be described by the soft-only CQM. Careful comparison of $N_{qp}$ with integrated $n_{ch}$ \auau\ centrality trends for more-peripheral collisions reveals substantial discrepancies (factor six difference in slopes below the sharp transition). The detailed evolution of \auau\ spectrum structure and angular correlations cannot be explained by a homogeneous model of soft production. The exact correspondence of jet-related correlated pair number and spectrum integrals based on a pQCD dijet frequency is also incompatible with the CQM.

A conjectured algebraic relation between the TCM and MB distributions invoked to reject that model is found to be invalid. The correct algebraic relation between the TCM for production centrality trends and MB distributions is established in the present study. The TCM description of MB data is accurate and provides new insights into detailed structure, including consequences of production fluctuations in central \aa\ collisions. In contrast, the Extreme Independent Model relating the CQM to MB distributions is shown to be misleading. The apparent agreement with data results from the insensitivity of the conventional semilog plotting format. The only real model test is the role of fluctuations in central collisions (shape of MB tail structure), which the CQM fails.

Can MB distributions on \et\ and $n_{ch}$ test hadron production models? The answer is yes provided several conditions are met: (i) The MB plotting format must make all information visually accessible, (ii) distortions from production fluctuations in more-central collisions must be understood, (iii) systematic uncertainties relating to centrality determination in more-peripheral collisions must be controlled, (iv) a proper linear-superposition reference extrapolated from \pp\ collisions must be established and (v) a mathematically correct TCM model of MB distributions must be defined to extract quantitative data.

That being said, analysis of MB distributions however accurate must compete with more-differential methods (spectra and correlations) that provide much more information about collision dynamics. Collision models have already been rigorously tested by a combination of highly-differential spectrum and correlation analysis. The TCM prevails as a general framework within which model details can be evaluated, and dijet production is an essential feature of any model of nuclear collisions.

We conclude that the TCM is not a misleading proxy for a more-legitimate soft model based on constituent quarks. Instead, the CQM is a proxy based on a single accidental relation and is falsified by a complex of differential analysis results described accurately by a TCM consistent with the pQCD description of dijets.

This work was supported in part by the Office of Science of the U.S.\ DOE under grant DE-FG03-97ER41020. 

\begin{appendix}

\section{TCM and MB distributions} \label{ppeerat} \label{etdata}
 
Figures~17-19 of Ref.~\cite{phenix} show conjectures about a relation between the TCM and MB data distributions. The figures show comparisons between MB data distributions on \et\ and the extreme cases of pure participant or \nn\ binary-collision scaling (Figs.~17 and 18) as discussed in Sec.~\ref{tcmcq}. Since both hypotheses fail to describe the data the TCM is rejected. But the descriptions fail not from problems with the TCM but from misapplication of  the TCM to MB distributions as discussed in Sec.~\ref{cqmtcm}. In this Appendix we derived the correct relation between the TCM and MB distributions on \et\ and $n_{ch}$.

\subsection{The TCM and MB distribution structure} \label{mbstruct}

Since \meet\ in Fig.~\ref{met} (right panel) is nearly constant with centrality  the power-law treatment of MB distributions on $n_{ch}$ in Ref.~\cite{powerlaw} is a good approximation to the present case on \et. We assume some acceptance $\Delta \eta$ with integrated total charge $n_{ch}$. Reference~\cite{powerlaw} notes that the MB distribution on $N_{part}$ is almost exactly a power law $\propto N_{part}^{-3/4}$ and is therefore approximately constant on $N_{part}^{1/4}$ within some bounded interval (rectangular). The structure of the measured MB distribution on $n_{ch}^{1/4}$ (deviations from the rectangular participant-scaling distribution) may directly reveal three aspects of hadron production: (a) fluctuations in \pp\ collisions, (b) fluctuations in central \aa\ collisions and (c) the parameter $x$ in  the TCM, measured respectively by the slopes on the left and right ends and in the central part of the MB distribution.

The data MB distribution can be predicted quantitatively from the TCM in the following way. We assume the usual TCM expression for the charge yield $(2/N_{part}) n_{ch} = n_{pp}[1 + x(\nu - 1)]$, where $n_{pp}$ is the charge yield for \pp\ (\nn) collisions. Then invoking Eq.~(\ref{jacobeq}) with power-law variables we obtain
\bea
\frac{d\sigma}{dn_{ch}^{1/4}} &=&\left[ \frac{dn_{ch}^{1/4}}{d(N_{part}/2)^{1/4}} \right]^{-1} \frac{d\sigma}{d(N_{part}/2)^{1/4}} ,
\eea
where the first factor is a Jacobian that can be derived from the basic TCM yield expression and the second factor is the rectangular MB distribution representing participant scaling. The MB distribution on $n_{ch}^{1/4}$ is then
\bea \label{tcmmb}
\frac{d\sigma}{dn_{ch}^{1/4}} &=& 
 \frac{[1 + x(\nu  -1)]^{3/4}}{1 + x(\nu  -1) +x\nu/3} \times
\\ \nonumber
&&
 \frac{d\sigma}{n_{pp}^{1/4}d(N_{part}/2)^{1/4}},
\eea
where we have invoked the approximation $\nu \approx (N_{part}/2)^{1/3}$~\cite{powerlaw}.
If $x$ is constant for all centralities the first factor (inverse Jacobian) is approximately a straight line on $\nu$ with negative slope ($\approx 1 -7x\nu/12$) reflecting some fractional contribution to $n_{ch}(N_{part})$ from \nn\ binary-collision scaling. The $x$ inferred from differential spectrum data increases by a factor six from peripheral ($x \approx 0.015$) to central ($x \approx 0.1$) \auau\ collisions~\cite{ppprd,hardspec,anomalous,jetspec}. The slope of the first factor then varies accordingly. Note that in Eq.~(A5) of Ref.~\cite{powerlaw} the quantity $\nu^{5/4}/3$ should be $\nu/3$ as in Eq.~(\ref{tcmmb}).

 \begin{figure}[h]
  \includegraphics[width=3.3in]{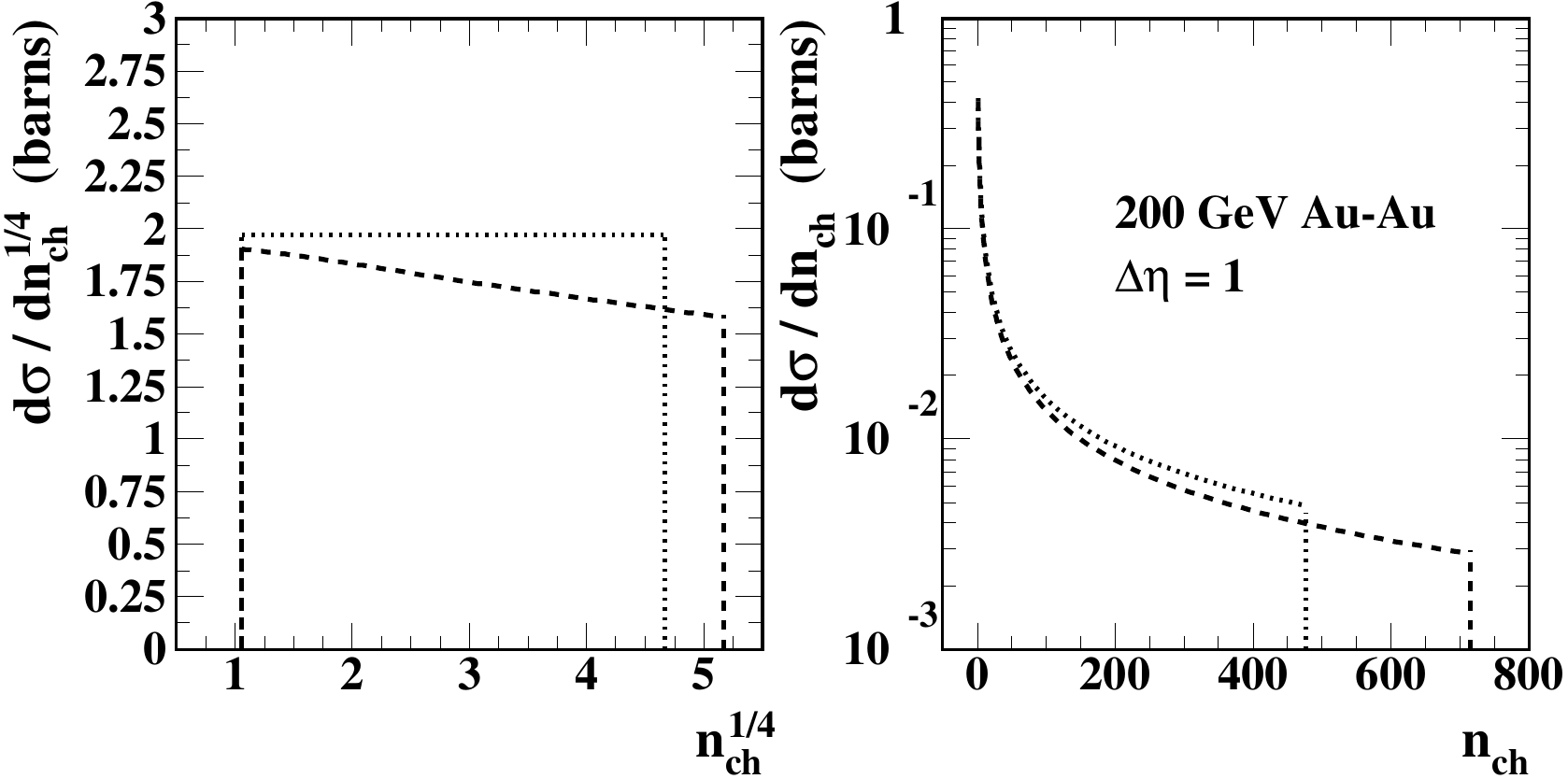}
\caption{\label{mb}
Left: MB distributions on $n_{ch}^{1/4}$ derived from Glauber Monte Carlo simulations~\cite{powerlaw} assuming participant scaling (dotted lines) and the conventional TCM with fixed $x = 0.1$ (dashed lines) according to Eq.~(\ref{tcmmb}).
Right:  The same trends plotted in the conventional MB semilog format on $n_{ch}$.
 } 
 \end{figure}

Figure~\ref{mb} (left panel) shows  MB distributions on $n_{ch}^{1/4}$ for (a) participant scaling (dotted lines, $x = 0$) and (b) the TCM describing more-central 200 GeV \auau\ collisions (dashed lines, fixed $x = 0.1$ for all centralities). Both distributions integrate to the \auau\ total cross section 7.2 barns. The sloping dashed line represents Eq.~(\ref{tcmmb}). In that plotting format the relation of distribution (b) to collision geometry and production mechanisms is easy to demonstrate (as discussed below). Figure~\ref{mb} (right panel) shows the same distributions in the conventional semilog plotting format, with Jacobian factor $dn^{1/4}/dn = 1/4 n^{3/4}$. Connections to the underlying particle production mechanisms and collision geometry are not visually accessible. The straightforward TCM implementation in Eq.~(\ref{tcmmb}) and Fig.~\ref{mb} is dramatically different from the conjectures in Figs.~17-19 of Ref.~\cite{phenix}.

\begin{figure}[h]
  \includegraphics[width=3.3in]{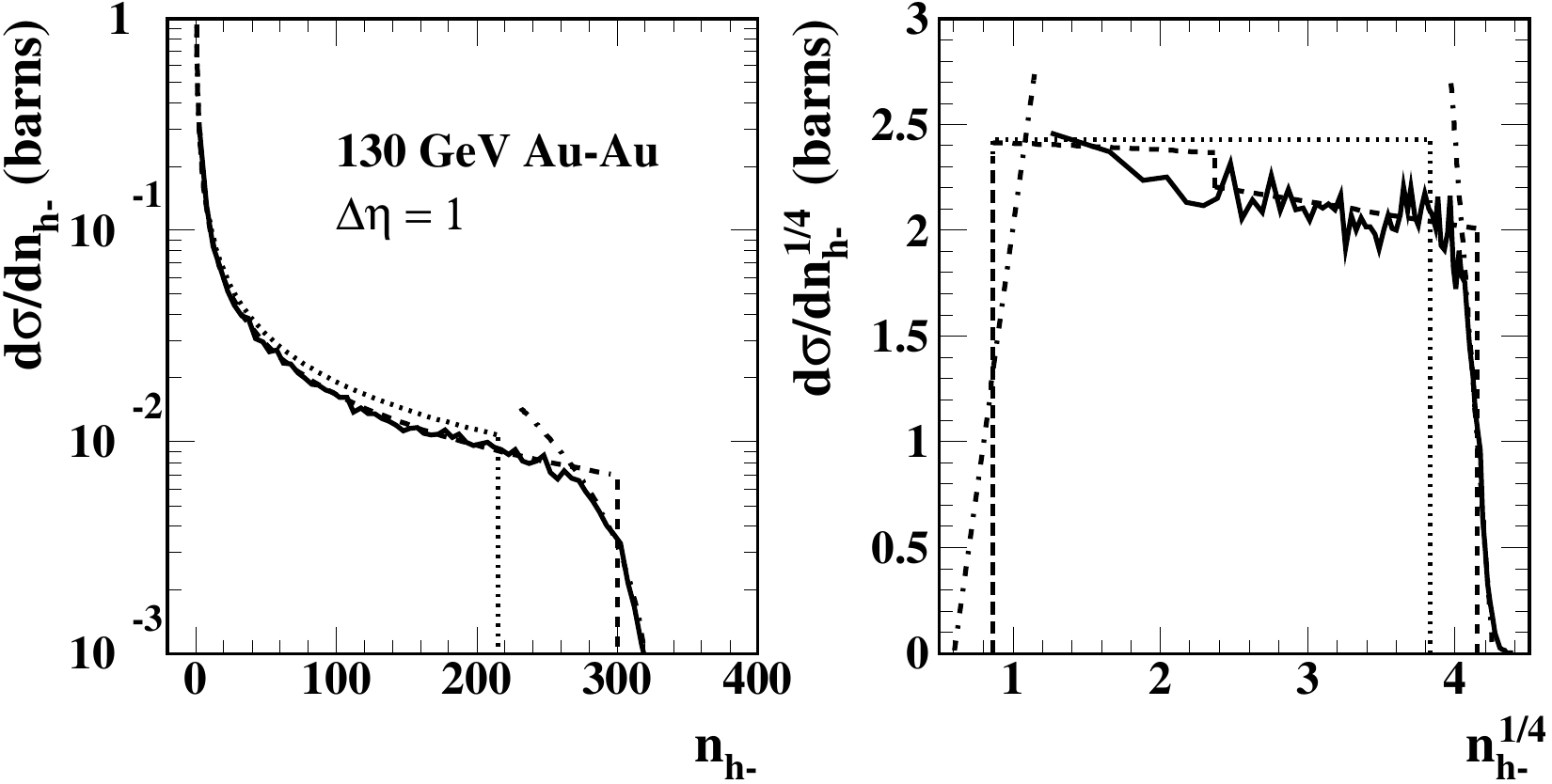}
\caption{\label{hmn}
Left: MB distribution of negative hadrons $h^{-}$ from 130 GeV \auau\ collisions (solid curve~\cite{hminus}) compared to power-law participant scaling (dotted curve) and the TCM with fixed $x = 0.08$ (dashed curve). The dash-dotted curve is discussed in the text.
Right: The same distributions plotted in power-law format. Information in the data is only the more-central mean slope (reflecting the TCM $x$ value for more-central collisions) and the slope at the terminus reflecting hadron production fluctuations in \auau\ collisions with $N_{part,max} = 382$. The step in the dashed curve at mid-centrality is a schematic model of  the ST (see the text).
 }  
 \end{figure}

Figure~\ref{hmn} (left panel) shows a comparison between STAR 130 GeV $h^-$ data (solid curve~\cite{hminus}) and the corresponding TCM of Eq.~(\ref{tcmmb}) (dashed curve) with constant $x = 0.08$ and $n_{pp} = 2.25/2$ (approximate $h^-$ $\eta$ density near $\eta = 0$ for 130 GeV). The dotted curve represents participant scaling. A similar comparison appears in Fig.~3 of Ref.~\cite{powerlaw}.  This semilog plotting format obscures essential data features relating to the TCM.

Figure~\ref{hmn} (right panel) shows the same distributions in the power-law format. The dotted lines represent participant scaling and the dashed lines (for more-central collisions) represent Eq.~(\ref{tcmmb}) with $x = 0.08$. Each distribution integrates to $\sigma_0 = 7.2$ barns. The model curves extend over the full fractional centrality interval $\sigma/\sigma_0 \in [0,1]$. The MB data and TCM agree to a few percent, well within systematic uncertainties. The TCM depends only on parameters $n_{pp} \equiv 2.25/2$ and $x = 0.08$ and otherwise was not fitted to the MB $h^-$ data. 
The right-hand dash-dotted line in the right panel is a good approximation to the tail structure in the left panel when transformed to that semilog format (dash-dotted curve).

The power-law format provides a precise estimate of the end-point multiplicity as the half-max point on the terminus (steeply falling part) of the MB distribution at the right end. That quantity is not accessible with the conventional semilog plotting format.   For these data the endpoint lies at $n_{h^-} = 4.15^4 = 300$. From the  corresponding \auau\ participant endpoint $N_{part}/2 = 191$  
we obtain $(2/N_{part}) n_{h^-} = 1.57$. Given $n_{h^-,pp} = 2.25/2 = 1.12$ for 130 GeV NSD \pp\ collisions we obtain
 $1+ x(\nu - 1) = 1.4$ at $\nu \approx 6$, implying $x \approx 0.08$. Thus, the terminus endpoint position accurately corresponds to TCM parameter $x$ and the slope of the dashed curve described by Eq.~(\ref{tcmmb}), at least for more-central  collisions.
 
Also included in the right panel is a schematic model of the ST observed at 62.4 and 200 GeV. The dashed curve for more-peripheral collisions follows the GLS trend described by Eq.~(\ref{tcmmb}) with $x = 0.02$. Extrapolated to central collisions the endpoint in that case would be $[1.1 \times (2.25/2) \times 191]^{1/4} = 3.92$ or $n_{h^-} = 236$. Near the ST at mid-centrality parameter $x$ increases from 0.02 to 0.08 (short vertical line), with endpoint $n_{h^-} = 300$ as noted.
 
From these comparisons we find that one learns little from MB distributions in the conventional semilog plotting format but a substantial amount from data in the power-law format. Relative to the power-law Glauber  model one can estimate \pp\ fluctuations, central \aa\ fluctuations and centrality evolution of TCM parameter $x$. The consequence of dijet production in \aa\ collisions (TCM hard component) is redistribution of the \aa\ cross section to larger integrated yields according to Eq.~(\ref{tcmmb}).

\subsection{Production fluctuations in central collisions}  \label{tailflucts}

Fluctuations in particle and \et/\pt\ production produce a distortion of MB distributions. Differences in detector angular acceptances can lead to corresponding differences in MB distributions. The terminus slope in the power-law format (e.g.\ Fig.~\ref{hmn}, right-hand dash-dotted line) represents production fluctuations in central collisions ($b \approx 0$). The terminus slope $m$ is proportional to $1/\sigma_n$ measuring charge multiplicity fluctuations (the variance depends on the angular acceptance). The mean slope of the right-hand terminus is  $|m| \approx 10$, and the Gaussian r.m.s.\ on $n_{h^-}^{1/4}$ is $\sigma_n = 2/\sqrt{2\pi} |m| \approx 0.08$. The corresponding fluctuation manifestation on $n_{h^-}$ is estimated by the dash-dotted curve in the left panel, with relative r.m.s.\  on $n_{h^-}$ $4 \times\sigma_n / 300^{1/4} \approx 0.08 $ or 8\% for charge-number correlations within acceptance $\Delta \eta = 1$. Production fluctuations in \pp\ collisions from the terminus sketched by the left dash-dotted line aren't accessible because these MB data do not extend low enough on $n_{h^-}^{1/4}$.

The terminus fluctuation structure in the MB distribution has major consequences for plots of production vs centrality. We can determine fluctuation effects in the latter using the exact correspondence between a power-law MB distribution and the corresponding plot of production on centrality.

\begin{figure}[h]
  \includegraphics[width=3.3in]{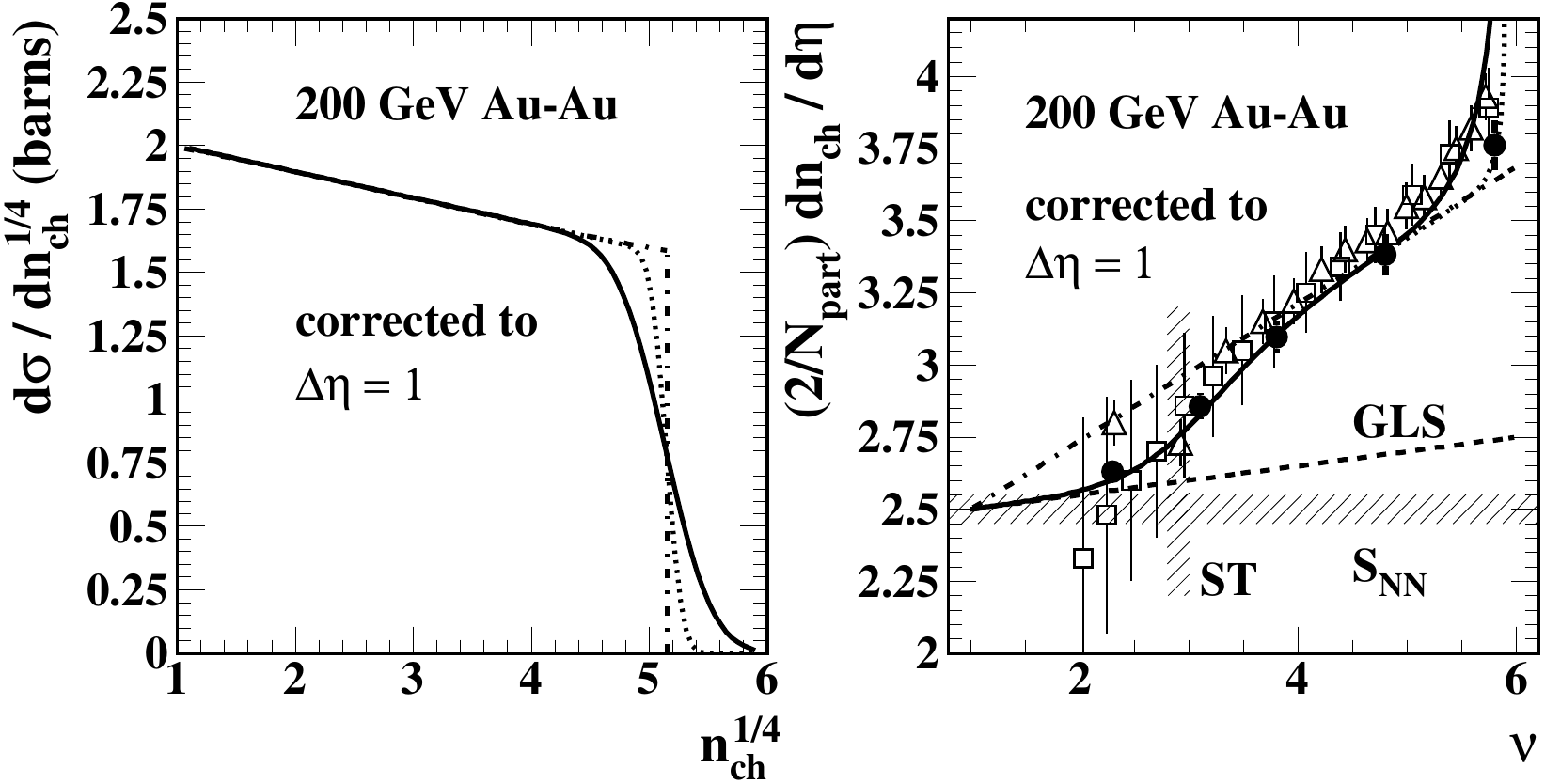}
\caption{\label{flucts}
Left: Models of MB frequency distributions on $n_{ch}$ in the power-law format including no fluctuations (dash-dotted curve), fluctuations for the STAR TPC acceptance (dotted curve) and for the fiducial PHENIX EMCal acceptance (solid curve), both corrected to the reference acceptance.
Right: Measured production centrality trends for STAR (solid dots)~\cite{hardspec} and  PHENIX (open squares) and ``RHIC average'' (open triangles)~\cite{phenixnch} compared to integrated model MB distributions from the left panel (same line styles). The large effect of angular acceptance on fluctuation distortions for more-central collisions is demonstrated.
 } 
 \end{figure}

Figure~\ref{flucts} (left panel) shows the TCM model for 200 GeV \auau\ collisions (dash-dotted lines) with $x = 0.095$. The dotted curve represents that curve folded with a Gaussian (approximated by an error function) having a fluctuation width corresponding to the STAR TPC acceptance (see Fig.~\ref{hmn} -- right panel). For the PHENIX EMCal acceptance the corresponding width is three times larger (solid curve). We transform MB fluctuation models to production centrality trends as follows. The running integral of a curve in the left panel yields $\sigma(n_{ch}^{1/4})$ vs $n_{ch}^{1/4}(\sigma)$. We divide $n_{ch}(\sigma)$ by $N_{part}(\sigma)/2$ and convert $\sigma/\sigma_0$ to $\nu$ using Eqs.~(\ref{glauber}) to obtain the required production centrality dependence $(2/N_{part}) dn_{ch}/d\eta$ vs $\nu$.

Figure~\ref{flucts} (right panel) shows the integration results as the solid and dotted curves plotted above $\nu = 4.5$. The solid curve below $\nu = 4.5$ represents the ST trend as derived from jet-related angular correlations and shown in Fig.~\ref{corresp} (right panel). Corresponding PHENIX and ``RHIC average'' production data are plotted as open squares and open triangles respectively~\cite{phenixnch} compared to STAR spectrum integrals (solid dots)~\cite{hardspec}.
This exercise demonstrates that the ten-times smaller PHENIX EMCal angular acceptance results in a substantial fluctuation distortion or bias for more-central collisions, equivalent to about 18\% relative fluctuations compared to 6\% for STAR within $\Delta \eta = 2$ . The consequence for STAR data is a single most-central point falling slightly above the fixed-$x$ TCM dash-dotted line.

Comparing the right panels of Figs.~\ref{hmn} and \ref{flucts} we observe that a MB distribution on produced quantity $X$ is more differential than the $X$ production centrality trend, but also includes substantial statistical noise. The running integral leading to the production centrality trend does filter the noise to better reveal the information. But the same noise reduction may be accomplished by rebinning the MB distribution (uniform bins in the power-law format). Both formats convey the same information.

\section{TCM vs Collision energy} \label{rhiclhc}

Reference~\cite{phenix} presents arguments that the TCM, including its hard component representing dijet production, is actually a proxy for soft production scaling with the number of constituent-quark participants. LHC hadron production data~\cite{alice} are invoked to argue that the centrality trend for hadron production from 200 GeV \auau\ is essentially the same as that from 2.76 TeV \pbpb\ collisions, although ``the jet cross section increases by a large factor.'' Thus, hadron production at both energies must be dominated by soft processes with negligible dijet contribution. However, the detailed structure of the data trends and the predicted energy evolution of the TCM actually support the TCM interpretation.

\subsection{Predicting the energy evolution of the TCM}

Two features of 2D angular correlations at RHIC have been observed to scale with collision energy as $\log(\sqrt{s_{NN}} / Q_0)$. For the nonjet quadrupole $Q_0 \approx 13.5$ GeV~\cite{davidhq}. For dijet production {\em per final-state hadron} in $p_t$ correlations  $Q_0 \approx 10$ GeV~\cite{ptedep}. A jet-production lower bound near $Q_0 \approx 10$ GeV is consistent with the observed jet-energy lower bound 2 - 3 GeV for parton fragmentation to charged hadrons~\cite{fragevo}. For jet-related number correlations we observe that $\log(\sqrt{s_{NN}} / \text{9 GeV})$ predicts factor 1.6 in Fig.~\ref{aacorr2} (left panel) relating dijet production per hadron in \auau\ collisions at 200 vs 62.4 GeV~\cite{anomalous} and predicts factor 2.2 relating equivalent dijet structure in \pp\ collisions at 200 GeV vs 7 TeV~\cite{cmsridge}. The corresponding factor relating 200 GeV to 2.76 TeV is 1.8. 

Hadron production in \pp\ collisions inferred from \yt\ spectrum $n_{ch}$ dependence is described in Sec.~\ref{ppmb}:  (a) small-$x$ gluons estimated by soft yield $n_s$ provide the common underlying degree of freedom in high-energy \nn\ collisions, and (b) the dijet production trend is indicated by hard yield $n_h \propto n_s^2$. If the relative density of small-$x$ gluons produced at a given \nn\ collision energy scales proportional to $\log(\sqrt{s}/Q_0)$ and $n_h/n_s \propto n_s$ then observed dijet energy trends are explained quantitatively. 

 The TCM centrality trend for 200 GeV \auau\ collisions is defined by $2.5(1+x(\nu - 1))$ with GLS $x \approx 0.02$ below the ST and modified-FF $x \approx$ 0.095 above the ST. According to the above argument the TCM for 2.76 TeV is predicted by scaling up the soft component (increased small-$x$ gluons) by factor 1.8 and the hard component {\em relative to the soft component} by another factor 1.8. The absolute increase in dijet production should then be factor $1.8^2 = 3.24$ (increased collisions of small-$x$ gluons).

\subsection{Charged-hadron production at two energies} \label{twoenergies}

\begin{figure}[h]
  \includegraphics[width=3.3in]{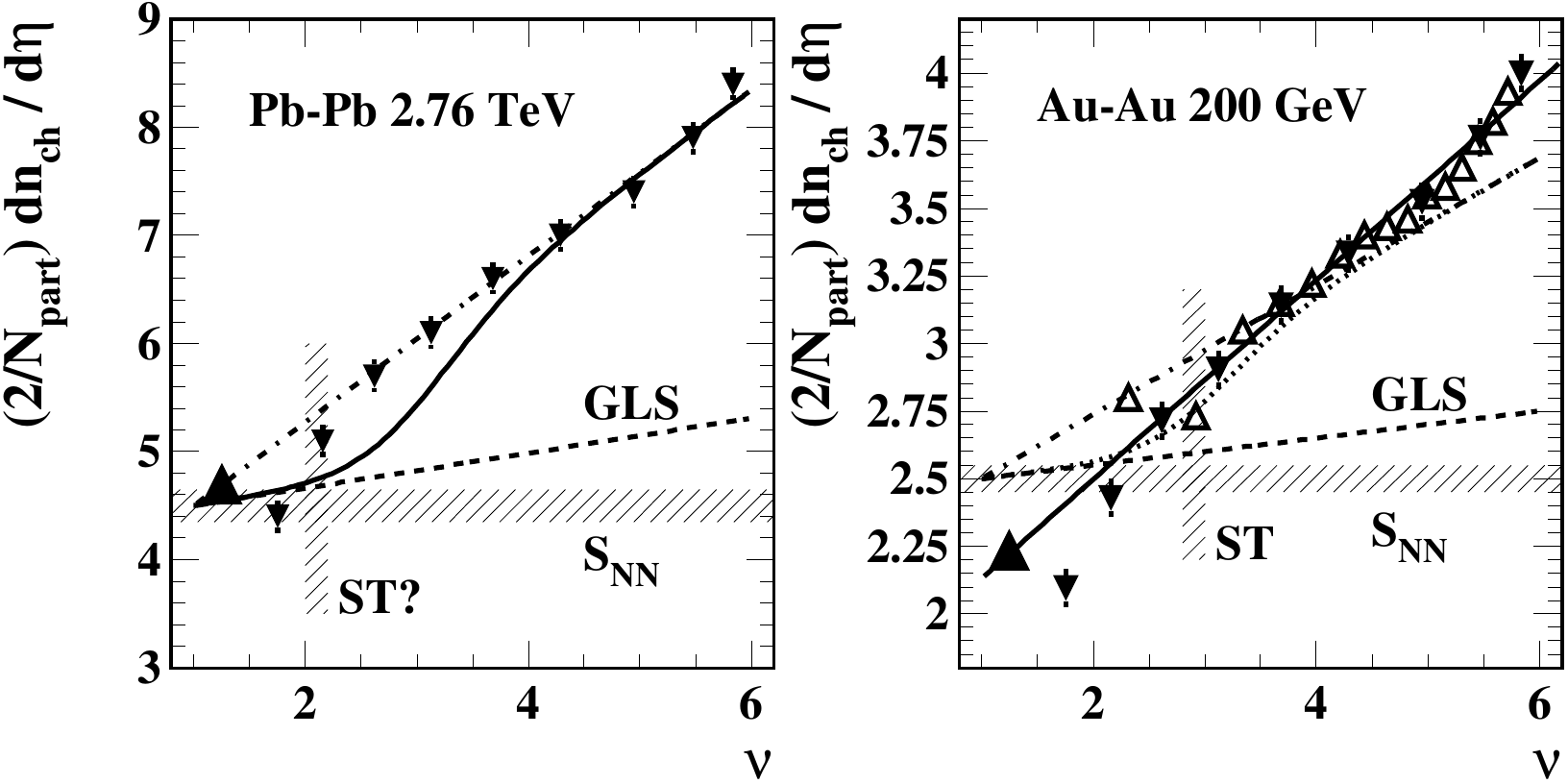}
\caption{\label{alice}
Left: Hadron production vs centrality for 2.76 TeV \pbpb\ collisions (inverted solid triangles~\cite{alice}) compared to TCM trends (dash-dotted and dashed lines) extrapolated from the 200 GeV TCM based on measured RHIC energy trends as described in the text. The upright solid triangle denotes \pp\ data.
Right: Hadron production vs centrality for 200 GeV \auau\ collisions  (open triangles~\cite{phenixnch}) compared to a TCM prediction from analysis of jet-related angular correlations (dotted curve)~\cite{jetspec}. The dash-dotted line indicates the conventional  200 GeV \auau\ TCM with fixed $x = 0.095$~\cite{kn}. The dashed line indicates a GLS extrapolation from \pp\ collisions with $x= 0.015$. 
The solid triangles and solid line are the solid triangles and dash-dotted line in the left panel scaled down by factor 2.1. The two line slopes differ by factor 1.55.
 } 
 \end{figure}

Figure~\ref{alice} (left panel) shows per-participant-pair charge production data from 2.76 TeV \pbpb\ collisions (inverted solid triangles) and \pp\ collisions (upright solid triangle)~\cite{alice}.  To accommodate the 6\% larger atomic number of lead the path length is scaled up by 2\% since $\nu \propto N_{part}^{1/3}$ is a good approximation. The TCM limiting cases are defined as described above by  $1.8\times 2.5(1+1.8 \times x(\nu - 1))$ with $x = 0.02$ (dashed line) and $x = 0.095$ (dash-dotted line). The solid curve is the 200 GeV centrality trend inferred from jet-related correlations (reflecting the ST) scaled up in the same way.  Given the relation to data the 2.76 GeV hadron production data are {\em predicted} by a TCM  based on the energy dependence of soft and hard production below 200 GeV.  The TCM was not adjusted to accommodate the 2.76 TeV data.

Figure~\ref{alice} (right panel) shows the ``RHIC average'' hadron production data appearing in Fig.~\ref{flucts} (open triangles, reported in Ref.~\cite{phenixnch}) compared to the 200 GeV TCM. The inverted solid triangles are the 2.76 TeV data in the left panel scaled down by factor 2.1.  The apparent similarity has been invoked to support the conclusion that particle production follows the same general trend at two very different energies and therefore must be due exclusively to a soft process, inconsistent with the TCM~\cite{phenix}. In that comparison the 200 GeV data were actually scaled up by factor 2.1 to match the 2.76 TeV data, but this reversed-scaling comparison is equivalent.

The claimed similarity of data trends at two different energies is interpreted to imply a common soft hadron-production mechanism with no significant dijets from hard processes, but the comparison is misleading. The scaled 2.76 TeV data actually strongly disagree with the 200 GeV production data from Ref.~\cite{hardspec} (dotted curve) and the 200 GeV TCM in several ways:
(i) The average slope of the scaled higher-energy data (solid line) is $1.8^2/2.1 = 1.55$ times larger than the STAR 200 GeV data trend and TCM (dash-dotted line). The difference greatly exceeds what is allowed by various systematic uncertainties.
(ii) The apparent agreement for more-central data results from the substantial fluctuation bias in the PHENIX data, whereas that bias is small for the ALICE acceptance. The unbiased limiting case is the TCM dash-dotted line.
(iii) The downscaled data for peripheral 2.76 TeV \pbpb\ and \pp\ collisions fall a factor 1.8/2.1 = 0.86 below the 200 GeV NSD value 2.5 and inferred $S_{NN}$, again well beyond data uncertainties.
In contrast, the 200 GeV TCM scaled appropriately for small-$x$ parton participants and parton-parton binary collisions {\em predicts} a TCM in the left panel that describes the 2.76 TeV data within their uncertainties. 
Thus, the recent LHC data strongly favor a TCM based on small-$x$ parton energy dependence extrapolated from RHIC data trends.

\subsection{Possible energy evolution of the sharp transition} \label{sharpenergy}

The sharp transition first observed for minimum-bias jet-related angular correlations in \auau\ collisions and indicated by the hatched box ST in Fig.~\ref{alice} (right panel)~\cite{anomalous} is a remarkable new feature of hadron production systematics. Below the ST hadron production (spectra and correlations) follows the GLS reference extrapolated from \pp\ collisions. Above the ST hadron production within the hard component increases dramatically, the increase attributed mainly to modified parton fragmentation~\cite{fragevo}. While fragment yields are suppressed at larger \pt\ as indicated by ratio $R_{AA}$ they are much more enhanced at lower \pt\  (the increase concealed by the properties of $R_{AA}$) but remain within the correlated jet structure and conserve the parton energy~\cite{hardspec,jetspec}. The result is an increase in hadron fragments {\em within intact jets} of up to a factor 6 and no reduction in dijet number~\cite{fragevo}.

It could be argued that the most significant information in Fig.~\ref{alice} (left panel) is the possible migration of the ST downward in centrality with increasing collision energy. In Fig.~\ref{corresp} (left panel) the ST at 200 GeV (right hatched box) corresponds to a dijet $\eta$ density per \auau\ collision (solid curve) of approximately unity. The dashed curve is the solid curve scaled up by factor $1.8^2 \approx 3.25$ corresponding to the predicted absolute increase in dijet yield from 200 GeV to 2.76 TeV.  The left hatched box indicates the centrality for the same dijet density per \auau\ collision given the increased dijet rate per \nn\ collision. Comparing the two panels of Fig.~\ref{alice} we find that the 2.76 TeV data are consistent with a shift in the ST downward from $\nu \approx 3$ to $\nu \approx 2$ consistent with the exercise in Fig.~\ref{corresp}.  That correspondence suggests that the ST and modified parton fragmentation may be determined by the dijet density within \aa\ collisions.

\section{Joint MB distributions} \label{joint}

MB distributions such as Fig.~2 of Ref.~\cite{phenix} are related to a system of joint, conditional and marginal distributions of cross-section $\sigma$ or event-number density on some measured kinematic quantity $X$ (e.g.\ $n_{ch}$ or $E_t$) and one or more possible \aa\ model parameters $N_x$ derived from the 3D space $(X,N_{x},N_{y})$  by projection. The 3D density distribution is typically tightly correlated (localized) as expressed by a locus of mean values $(\bar X,\bar N_x,\bar N_y)$ realized for instance by Eq.~(\ref{tcmaa}) for $N_{part}$, $N_{bin}$ and $X = n_{ch}$. Projections onto subspaces and running integrals within subspaces are used to determine the locus of mean values and detailed features of projected MB distributions.

It is useful to establish an algebraic context for such projections, for example based on the integrated \et\ within some angular acceptance vs a model parameter $N_x$ such as $N_{part}/2$ or $N_{qp}$. The 2D joint MB distribution $F(E_t,N_x)$ may be normalized to unity (joint probability distribution) or to \aa\ total cross section $\sigma_0$. Applying the chain rule for joint probabilities we obtain
\bea
F(E_t,N_x) &=& \hat G(E_t|N_x) F(N_x)
\\ \nonumber
& =& \hat H(N_x|E_t) F(E_t),
\eea
where for instance $ \hat G(E_t|N_x)$ is the unit-normal conditional probability density distribution on \et\ given a specific $N_x$ condition, and $F(E_t)$ is the marginal projection of 2D $F(E_t,N_x)$ onto $E_t$ with
\bea \label{project}
F(E_t) &=& \int dN_x\, \hat G(E_t|N_x) F(N_x).
\eea
Equation~(\ref{project}) describes Fig.~2 of Ref.~\cite{phenix} where $ \hat G(E_t|N_x)$ represents the individual lower curves and $F(E_t)$ is the upper solid curve. In Eq.~(7) of Ref.~\cite{phenix} $\sigma_{BA}\, w_n \leftrightarrow F(N_x)$ and $P_n(E_t) \leftrightarrow  \hat G(E_t|N_x)$. For generality we treat $F(N_x)$ as a density on  continuous $N_x$ which could be a sum of delta functions at integer values.

Each conditional distribution $ \hat G(E_t|N_x)$ is a density on \et\ peaked near some conditional mean value $\bar E_t(N_x)$. On the other hand, for a given \et\ the ensemble of $ \hat G(E_t|N_x)$ can be seen as a distribution on $N_x$ peaked near some mean value $\bar N_x(E_t)$. The conditional distribution in the argument of the integral in Eq.~(\ref{project}) can be converted to a density on $N_x$ with a suitable Jacobian $J(E_t,N_x)$. Treating $J(E_t,N_x)\hat G(E_t|N_x)$ as a weight function on $N_x$ the integral in Eq.~(\ref{project}) can be interpreted as a weighted average of $F(N_x)/J(E_t,N_x)$ at given \et\ approximated by that ratio evaluated at $\bar N_x(E_t)$. With $J(E_t,N_x) = dE_t/dN_x$ we then obtain
\bea \label{jacobeq}
F(E_t) &=& [ dE_t/dN_x ]_{\bar N_x}^{-1}\,F[\bar N_x(E_t)],
\eea
where the Jacobian $dE_t/dN_x$ is a variable quantity derived from the locus (curve) of mean values $\bar E_t(N_x)$. Analysis of MB distributions indicates that in cases relevant to \aa\ collision geometry the details of conditional distributions $ \hat G(E_t|N_x)$, other than the conditional mean values $\bar E_t(N_x)$, are not relevant except at the ends (termini) of the $F(N_x)$ distribution~\cite{powerlaw}.

\end{appendix}



\begin{thebibliography}{99}

\bibitem{phenix}  S.~S.~Adler {\it et al.}  (PHENIX Collaboration),
  arXiv:1312.667.

\bibitem{voloshin}   S.~Eremin and S.~Voloshin,
  Phys.\ Rev.\ C {\bf 67}, 064905 (2003).

\bibitem{ppprd} J.~Adams {\it et al.}  (STAR Collaboration),
  Phys.\ Rev.\  D {\bf 74}, 032006 (2006).

\bibitem{porter2} R.~J.~Porter and T.~A.~Trainor  (STAR Collaboration),
  J.\ Phys.\ Conf.\ Ser.\  {\bf 27}, 98 (2005).

\bibitem{porter3}  R.~J.~Porter and T.~A.~Trainor  (STAR Collaboration),
  PoS {\bf CFRNC2006}, 004 (2006).

\bibitem{ua1} C.~Albajar {\it et al.}  (UA1 Collaboration),
  Nucl.\ Phys.\  B {\bf 309}, 405 (1988).

\bibitem{aleph}  D.~Buskulic {\it et al.}  (ALEPH Collaboration),
  Z.\ Phys.\  C {\bf 55}, 209 (1992).

\bibitem{opal} M.~Z. Akrawy {et al.}  (OPAL Collaboration)
  {Phys. Lett.} B, {\bf 247}, 617 (1990).

\bibitem{borg}   N.~Borghini and U.~A.~Wiedemann,
  hep-ph/0506218.

\bibitem{phobostcm} B.~B.~Back {\it et al.}  (PHOBOS Collaboration),
  Phys.\ Rev.\ C {\bf 65}, 061901 (2002).

\bibitem{fragevo} T.~A.~Trainor,
  Phys.\ Rev.\ C {\bf 80}, 044901 (2009).

\bibitem{kn}  D.~Kharzeev and M.~Nardi,
  Phys.\ Lett.\  B {\bf 507}, 121 (2001).

\bibitem{hardspec}  T.~A.~Trainor,
  Int.\ J.\ Mod.\ Phys.\  E {\bf 17}, 1499 (2008).
 
\bibitem{valons}  R.~C.~Hwa and M.~S.~Zahir,
  Phys.\ Rev.\ D {\bf 23}, 2539 (1981).

\bibitem{anomalous}  G.\ Agakishiev, {\it et al.} (STAR Collaboration),
  Phys.\ Rev.\ C {\bf 86}, 064902 (2012).

\bibitem{axialci}   J.~Adams {\it et al.}  (STAR Collaboration),
  Phys.\ Rev.\  C {\bf 73}, 064907 (2006).

\bibitem{inverse} T.~A.~Trainor, R.~J.~Porter and D.~J.~Prindle,
  J.\ Phys.\ G {\bf 31}, 809 (2005).

\bibitem{davidhq}  D.~T.~Kettler  (STAR collaboration),
  Eur.\ Phys.\ J.\  C {\bf 62}, 175 (2009).

\bibitem{davidhq2}  D.~Kettler ( STAR Collaboration),
  J.\ Phys.\ Conf.\ Ser.\  {\bf 270}, 012058 (2011).

\bibitem{ptfluct} J. Adams {\it et al.} (STAR Collaboration), 
Phys. Rev. C {\bf 71}, 064906 (2005). 

\bibitem{ptscale} J. Adams {\it et al.} (STAR Collaboration),
J. Phys. G {\bf 32}, L37 (2006).

\bibitem{ptedep}  J.~Adams {\it et al.}  (STAR Collaboration),
  J.\ Phys.\ G {\bf 33}, 451 (2007).

\bibitem{powerlaw}   T.\,A.~Trainor and D.\,J.~Prindle, 
 hep-ph/0411217.

\bibitem{eeprd}   T.~A.~Trainor and D.~T.~Kettler,
  Phys.\ Rev.\ D {\bf 74}, 034012 (2006).

\bibitem{hijing} X.-N. Wang,  Phys. Rev. D {\bf 46}, R1900 (1992); 
X.-N.~Wang and M.~Gyulassy,
  Phys.\ Rev.\  D {\bf 44}, 3501 (1991).

\bibitem{pythia}  T. Sj\"ostrand and M. van Zijl, Phys. Rev. D {\bf 36}, 2019 (1987);
T.~Sj\"ostrand,
Comput.\ Phys.\ Commun.\  {\bf 82}, 74 (1994); 
T.~Sj\"ostrand, L.~L\"onnblad, S.~Mrenna and P.~Skands,
hep-ph/0308153.

\bibitem{herwig}  M.~Bahr, S.~Gieseke, M.~A.~Gigg, D.~Grellscheid, K.~Hamilton, O.~Latunde-Dada, S.~Platzer and P.~Richardson {\it et al.},
  Eur.\ Phys.\ J.\ C {\bf 58}, 639 (2008).

\bibitem{elizabeth} E.~W.~Oldag (STAR Collaboration),
  J.\ Phys.\ Conf.\ Ser.\  {\bf 446}, 012023 (2013).

\bibitem{axialcd} J.~Adams {\it et al.}  (STAR Collaboration),
  Phys.\ Lett.\  B {\bf 634}, 347 (2006).

\bibitem{jetspec}   T.~A.~Trainor and D.~T.~Kettler,
  Phys.\ Rev.\ C {\bf 83}, 034903 (2011).

\bibitem{pptheory}  T.~A.~Trainor,
Phys.\ Rev.\ D {\bf 87}, 054005 (2013).

\bibitem{lphd} Ya.~I.~Azimov, Yu.~L.~Dokshitzer, V.~A.~Khoze, S.~I.~Troyan, Z. Phys. C {\bf 27}, 65 (1985),  Z. Phys. C {\bf 31}, 213 (1986).

\bibitem{hijscale} Q.\,J.~Liu, D.\,J.~Prindle and T.\,A.~Trainor, 
Phys. Lett. B {\bf 632}, 197 (2006). 

\bibitem{glasma1}  T.~A.~Trainor,
  J.\ Phys.\ G {\bf 39}, 095102 (2012).

\bibitem{lund}  B.~Andersson,
  Camb.\ Monogr.\ Part.\ Phys.\ Nucl.\ Phys.\ Cosmol.\  {\bf 7}, 1 (1997).

\bibitem{hminus} C.~Adler {\it et al.}  (STAR Collaboration),
  Phys.\ Rev.\ Lett.\  {\bf 87}, 112303 (2001).

\bibitem{phenixnch}  S.~S.~Adler {\it et al.}  (PHENIX Collaboration),
  Phys.\ Rev.\ C {\bf 71}, 034908 (2005),
  [Erratum-ibid.\ C {\bf 71}, 049901 (2005)].

\bibitem{alice}  K.~Aamodt {\it et al.}  (ALICE Collaboration),
  Phys.\ Rev.\ Lett.\  {\bf 106}, 032301 (2011).

\bibitem{cmsridge}  T.~A.~Trainor and D.~T.~Kettler,
  Phys.\ Rev.\ C {\bf 84}, 024910 (2011).

\end{thebibliography}
\end{document}